\documentclass[12pt]{article} 
\usepackage{epsfig}
\usepackage{a4}
\usepackage{latexsym}
\usepackage{cite}

\usepackage{pslatex}
\usepackage[latin1]{inputenc}
\usepackage[T1]{fontenc}

\renewcommand{\theequation}{\thesection.\arabic{equation}}
\newcommand{\gsim}{\raisebox{-0.07cm}{$\:\:\stackrel{>}{{\scriptstyle
 \sim}}\:\: $} }
\newcommand{\lsim}{\raisebox{-0.07cm}{$\:\:\stackrel{<}{{\scriptstyle
 \sim}}\:\: $} }
\newcommand{\eqLL}{\raisebox{-0.07cm}{$\;\stackrel{{\rm LL}}{=}\;$} }

\newcommand{\hspn}{{\hspace{-5mm}}}

\newcommand{\beq}{\begin{equation}}
\newcommand{\eeq}{\end{equation}}
\newcommand{\bea}{\begin{eqnarray}}
\newcommand{\eea}{\end{eqnarray}}
\newcommand{\nn}{\nonumber}
\newcommand{\nin}{\noindent}
\newcommand{\MSb}{$\overline{\mbox{MS}}$}
\newcommand{\as}{\alpha_{\rm s}}
\newcommand{\ass}{\alpha_{\rm s}^{\:\!2}}
\newcommand{\asth}{\alpha_{\rm s}^{\:\!3}}
\newcommand{\asfo}{\alpha_{\rm s}^{\:\!4}}
\newcommand{\ar}{a_{\rm s}}
\newcommand{\ars}{a_{\rm s}^{\,2}}
\newcommand{\art}{a_{\rm s}^{\,3}}
\newcommand{\ra}{\rightarrow}
\newcommand{\ep}{\epsilon}

\def\frct#1#2{\mbox{\large{$\frac{#1}{#2}\:\!$}}}

\begin{document}

\setlength{\parskip}{0.2cm}
\setlength{\baselineskip}{0.525cm}

\def\Qs{Q^{\:\! 2}}
\def\mus{\mu^{\:\! 2}}
\def\ca{{C_A}}
\def\cas{{C^{\: 2}_A}}
\def\cath{{C^{\: 3}_A}}
\def\cafo{{C^{\: 4}_A}}
\def\cafi{{C^{\: 5}_A}}
\def\cf{{C_F}}
\def\cfs{{C^{\: 2}_F}}
\def\cfth{{C^{\: 3}_F}}
\def\nf{{n^{}_{\! f}}}
\def\nfs{{n^{\,2}_{\! f}}}
\def\nft{{n^{\,3}_{\! f}}}
\def\nffo{{n^{\,4}_{\! f}}}

\def\z#1{{\zeta_{#1}}}
\def\zs#1{{\zeta_2^{\,#1}}}


\begin{titlepage}

\noindent
LTH 952 \hspace*{\fill} July 2012\\
LPN12-083 \\
\vspace{2.5cm}
\begin{center}
\Large
{\bf Resummed small-x and first-moment evolution\\[1mm]
of fragmentation functions in perturbative QCD}\\
\vspace{1.5cm}
\large
C.-H. Kom$^{\, a}$, A. Vogt$^{\, a}$ and K. Yeats$^{\, b}$\\
\vspace{1cm}
\normalsize
\vspace{0.5cm}
{\it $^a$Department of Mathematical Sciences, University of Liverpool \\[0.3mm]
Liverpool L69 3BX, United Kingdom}\\
\vspace{0.3cm}
{\it $^b$Department of Mathematics, Simon Fraser University \\[0.3mm]
University Drive, Burnaby, BC, V5A 1S6, Canada}\\
\vfill
\large
{\bf Abstract}
\vspace{-0.2cm}
\end{center}
We study the splitting functions for the evolution of fragmentation 
distributions and the coefficient functions for single-hadron production in 
semi-inclusive $e^+e^-$ annihilation in massless perturbative QCD for small 
values of the momentum fraction and scaling variable $x$, where their 
fixed-order approximations are completely destabilized by huge 
double logarithms of the form $\as^{\,n\,} x^{\,-1} \ln^{\,2\:\!n-a\!}x$.
Complete analytic all-order expressions in Mellin-$N$ space are presented for 
the resummation of these terms at the next-to-next-to-leading logarithmic 
accuracy.
The poles for the first moments, related to the evolution of hadron 
multiplicities, and the small-$x$ instabilities of the next-to-leading order
splitting and coefficient functions are removed by this resummation, which 
leads to an oscillatory small-$x$ behaviour and functions that can be used at 
$N\!=\!1$ and down to extremely small values of $x$.
First steps are presented towards extending these results to the higher 
accuracy required for an all-$x$ combination with the state-of-the-art 
next-to-next-to-leading order large-$x$ results.

\vspace{1cm}
\end{titlepage}

\section{Introduction}

\vspace{-2mm}
The fragmentation distributions, or parton fragmentation functions, 
$\,D_{\! p}^{\,h}(x,\mus)$ encode the probability of a final-state parton $p$ 
in a hard scattering process to end up in (or fragment into) a hadron $h$ which
carries a fraction $x$ of the momentum of the parent (anti-)$\,$quark $q$ or 
gluon $g$.
Like their even more important initial-state counterparts, the parton
distributions of hadrons $f_{\! p}^{\,h}(x,\mus)$, these quantities include 
long-distance information and are thus not calculable in perturbative
Quantum Chromodynamics (QCD).
Their dependence on the fragmentation (final-state mass factorization) scale 
$\mu$, to be chosen of the order of a physical hard scale in the scattering 
process under consideration, is however calculable via the renormalization-%
group evolution equations
\beq
\label{Devol}
  \frac{\partial}{\partial \ln \mus} \; D_{i}^{\,h} (x,\mus) \:\: = \:\:
  \sum_{j\,=\,\rm q,\,g} \big[\, P^{\,T}_{\!ji}\!\left(\as (\mus)\right) 
  \otimes D_{\!j}^{\,h}(\mus) \,\big](x) \; .
\eeq
Here $\otimes$ denotes the standard Mellin convolution in the first arguments, 
\beq
\label{Mconv}
  \left[\, f_1^{}(\mus) \otimes f_2^{}(\mus) \right](x) \:\: \equiv \:\:
  \int_x^1 \frac{dy}{y} \; f_1^{}(y,\:\mus)\, 
  f_2^{} \Big(\, \frac{x}{y},\:\mus \Big) \; ,
\eeq 
which is reduced to a simple product by the transformation to Mellin moments,
\beq
\label{Mtrf}
  f_i^{}(N,\:\mus) \:\: = \:\: \int_0^1\! dx\; x^{\,N-1} f_i^{}(x,\:\mus) \; .
\eeq
The final-state (`timelike') splitting functions $P^{\,T}_{\!ji}$ in 
Eq.~(\ref{Devol}) admit an expansion in power of the renormalized strong 
coupling constant, here normalized as $\ar \equiv \as(\mus)/(4\pi)$, 
\bea
\label{Pexp}
 P^{\,T}_{\! ji}\!\left(x,\as (\mus)\right) \:\:  = \:\:
 \ar\, P^{\,T(0)}_{\! ji}(x) \,+\: \ars P^{\,T(1)}_{\! ji}(x) 
 \,+\: \art P^{\,T(2)}_{\! ji}(x) \,+\; \ldots \:\; ,
\eea
where we have, without loss of information, identified the coupling-constant
renormalization and mass-factorization scales.

A benchmark process for parton fragmentation is semi-inclusive $e^+e^-$ 
annihilation (SIA),
\beq
  e^+\,e^- \;\ra\;\, \gamma^{\,\ast}\,,\:Z\,,\:H \;\ra\; h + X \; ,
\eeq
which is closely related to deep-inelastic scattering (DIS), $e\:\!h \,\ra\, e 
+ X$, via the exchange of a virtual photon $\gamma^{\,\ast}$, Z-boson or Higgs 
particle. In the former (latter) case the four-momentum $q$ of the exchanged 
boson is timelike, $q^{\,2} > 0$ (spacelike, $q^{\,2} < 0$). $X$ stands for all
hadronic final states allowed  by quantum number conservation.
The cross section for vector-boson exchange can be written as \cite{NasWeb94}
\beq
\label{d2sigma}
  \frct{1}{\sigma_0^{}}\, \frct{d^{\:\!2} \sigma}{dx\; d\!\cos \theta}
  \:\: = \:\:  \frct{3}{8}\, (1 + \cos^2 \theta) \; F_T^{\:\! h}(x,\Qs)
         \,+\, \frct{3}{4}\, \sin^2 \theta \; F_L^{\:\! h}(x,\Qs) 
         \,+\, \frct{3}{4}\, \cos  \theta \;F_{\:\!\!A}^{\:\!h}(x,\Qs) \; ,
\eeq
where $\theta$ is the angle, in the center-of-mass (CM) frame, between the 
incoming electron beam and the hadron $h$ observed with four-momentum $p$, and 
the scaling variable is given by $x= 2pq/\Qs$ where $\Qs\equiv q^{\,2}$ in SIA. 
The dimensionless transverse ($T$) and longitudinal ($L$) fragmentation 
functions in Eq.~(\ref{d2sigma}), and the total fragmentation function 
$F_I^{\:\! h} = F_T^{\:\! h} + F_L^{\:\! h}$ 
obtained by integrating Eq.~(\ref{d2sigma}) over $\theta$, have been measured 
at LEP and earlier $e^+e^-$ colliders, see Ref.~\cite{PDG2012} for a general 
overview.
The interference of vector and axial-vector contributions leads to the 
parity-violating $\cos \theta$ term. The corresponding fragmentation function 
$F_{\!A}$ does not receive $1/x$ contributions and will not be considered in 
this article; it should be ignored when Eq.~(\ref{d2sigma}) is referred to 
below.

Up to corrections suppressed by powers of the CM energy $\sqrt{s} = Q$,  
the fragmentation functions can be expressed in terms of the fragmentation 
distributions and coefficient functions $C_{a,i\,}$,
\beq
\label{Fconv}
  F_a^{\:\! h}(x,\Qs) \:\: = \:\: \sum_{i \,=\, \rm q,\,g} 
  [\, C_{a,i} (\Qs) \otimes D_i^{\,h} (\Qs) ] (x) \; ,
\eeq
which are short-distance quantities and calculable in perturbation theory,
\beq
\label{Cexp}
 C_{a,i}(x,\as) \:\:=\:\: 
 (\delta_{aT}\, \delta_{iq} \,+\, \delta_{a\phi}\, \delta_{ig})\, \delta(1-x)
 \,+\, \ar\, c_{a,i}^{(1)}(x) \,+\, \ars c_{a,i}^{(2)}(x) \,+\: \ldots \;\; .
\eeq
Here we have identified, again without loss of information, the scale $\mu$ in 
Eq.~(\ref{Devol}) with the physical scale $Q$, and we have suppressed 
electroweak charge factors. 
Besides the quantities in Eq.~(\ref{d2sigma}), we have included the 
fragmentation functions $F_\phi$ for the exchange of a scalar $\phi$ coupling 
(like the Higgs-boson in the limit of a heavy top quark and $\nf$ massless 
flavours \cite{HGGeff}) directly only to gluons via $\phi\,G_{\mu\nu}^{\,a} 
G_a^{\:\mu\nu}\!$, where $G^{\,a}_{\mu\nu}$ denotes the gluon field strength 
tensor.

The splitting functions (\ref{Pexp}) and coefficient functions (\ref{Cexp}) 
are known (with a minor caveat in the former case which is not relevant in the 
present context) to order $\asth$ and $\ass$, respectively, see Refs.\
\cite{FPplb,RvN-FTL,MMV06,MMoch06,MV2,AMV1} and references therein. 
These results provide the next-to-next-to-leading order (NNLO) approximation 
of (renormalization-group improved) perturbative QCD except for $F_L$, where 
the third-order contributions to Eq.~(\ref{Cexp}) would also be needed.
All these quantities exhibit, in contrast to their initial-state and DIS
counterparts \cite{ZvN-F2L,MVerm99,MVV34}, a double-logarithmic enhancement at 
small $x$, i.e., the contributions at order $\as^{\:\!n}$ are enhanced by terms
of the form $x^{\,-1}\ln^{\,2n-a} x$ (the minimal offset $a$ depends on the 
quantity under consideration) which correspond to poles $\as^{\,n}/
(N-1)^{\,2n+1-a}$ after performing the Mellin transformation (\ref{Mtrf}).
These terms spoil the convergence of the expansions (\ref{Pexp}) and 
(\ref{Cexp}) already at $x \lsim 10^{\,-2}$ \cite{MV2,AMV1} and, obviously, 
preclude describing particle numbers (multiplicities) given by the first 
moments, $N=1$, of the fragmentation distributions $\,D_{\! p}^{\,h}$.

At leading and next-to-leading logarithmic (LL and NLL) accuracy, these issues 
were addressed long ago, see Refs.~\cite{LLxto0,Mueller83,GMuell85}, by 
showing that these small-$x$ contributions can be calculated to all orders. 
For example, the leading logarithms ($a = 2$) of $P^{\,T}_{\!\rm gg}$ can be 
resummed to yield
\beq
\label{PLLsum}
  P_{\rm gg}^{\,T}(N,\as) \:\:\eqLL\:\:
  \frct{1}{4}\: (N\!-\!1) \Big\{ \!
    \left( 1 + \frct{32\:\! \ca \ar}{(N-1)^2}  \right)^{1/2}\! - 1
  \Big\} 
\eeq
with $\ca = n_{\rm colours} = 3$ in QCD, which leads to a contribution
proportional to $\sqrt{\as}$ at $N=1$. 
More recently there has been renewed interest in the all-$x$ and $N=1$ 
evolution of (parton) fragmentation functions, see, e.g., 
Refs.~\cite{AKKO05,ABKK11,Bol-DIS12}. In particular, a new method has been 
developed in Ref.~\cite{AV2011} for carrying out the small-$x$ resummation up 
to the third logarithms in the standard \MSb\ factorization scheme not adopted
in Ref.~\cite{Mueller83} (see Ref.~\cite{ABKK11} for a more detailed discussion
of this issue).

This article builds upon and extends the results of Ref.~\cite{AV2011} which 
were mostly presented in terms of perturbative coefficients to order 
$\as^{16}$. 
While this is sufficient for collider applications down to $x\simeq 10^{\,-4}$,
it does neither fix the first moments nor cover the vastly wider $x$-range 
relevant for fragmentation processes induced by ultra-high energy cosmic rays, 
see, e.g., Ref.~\cite{UHECR}. 
Here, we provide analytic resummed small-$x$ expressions at next-to-next-to-%
leading logarithmic (NNLL) accuracy which facilitate an all-$x$ `NLO $+$ 
resummed' evolution of the fragmentation functions, and derive the third terms 
in the resulting $N=1$ expansion in powers of $\sqrt{\as}$. 
These results include the case of $F_L$, which was dealt with only at NLL level
in Ref.~\cite{AV2011}. Furthermore we use the approach of 
Dokshitzer, Marchesini and Salam (DMS) in Ref.~\cite{DMS05},
which relates the evolution of flavour non-singlet fragmentation and parton
distributions, to provide an alternative derivation of the results for the  
`non-singlet' part of $P_{\!\rm gg}^{\,T}$ and to extend its resummation to the 
fifth logarithms and the $\sqrt{\as}^{\,5}$ contributions at~$N=1$ in a manner
outlined already in Ref.~\cite{March06}, see also  Ref.~\cite{DMarch07}.

\vspace{-2mm}
\setcounter{equation}{0}
\section{Formalism of the resummation}

\vspace{-2mm}
Before we present our results, we briefly recall the formalism for deriving
the resummation, which is based on the mass-factorization relations and the
structure of the unfactorized expressions 
\beq
\label{FThatFact}
  \widehat{F}_{a,k} \:\:=\:\! \widetilde{C}_{a,\,i} \,\otimes\,
  Z_{ik}^{\:\!T} \;
 \quad \mbox{ for } \quad a \:=\: T,\,\phi,\,L 
 \quad \mbox{ and } \quad k \:=\: {\rm q,\, g}
\eeq
in dimensional regularization (we use $D = 4-2\:\!\ep$).
The functions $\widetilde{C}_{a,\,i}$ are given by Taylor series in $\ep$, 
with the $\ep^{\:\!k}$ terms including $k$ more powers in $\ln x$ than the 
4-dimensional coefficient functions~(\ref{Cexp}). 
The transition matrix $Z^{\,T}$ consist of only negative powers of $\ep$ and
can be written in terms of the splitting functions (\ref{Pexp}) and the
expansion coefficients $\beta_n$ of the beta function of QCD, $\beta(\as) \,=\,
-\beta_0^{}\,\ar^2 \,-\: \ldots$ with $\beta_0^{} \,=\, \frac{11}{3}\:\ca 
- \frac{2}{3}\:\nf$.
This dependence can be summarized as
\beq
\label{ZofPn}
  \ar^{\,n}\,\ep^{\,-n}:\: P_{\:\!0}^{\,T},\: \beta_0^{} \; ; \quad
  \ar^{\,n}\,\ep^{\,-n+1}: \:\ldots\,+\, P_{\:\!1}^{\,T},\: \beta_1^{} \;; \;\;
  \dots \; ; \quad
  \ar^{\,n}\,\ep^{\,-1}:\: P_{n-1}^{\,T} \; .
\eeq
Hence fixed-order knowledge at N$^m$LO (i.e., of the splitting functions to
$P_m$ and the corresponding coefficient functions) fixes the first $m\!+\!1$
coefficients in the $\ep$-expansion of $\widehat{F}_{a,k}$ at all orders
in~$\as$.

The small-$x$ expansions of $\widehat{F}_{T,\,\phi}$ 
(the corresponding relation for $F_L$ is slightly different) read
\beq
\label{FThatLogs}
  \widehat{F} \big|_{\ar^{\,n}\,\ep^{\,-n+\ell}} \;=\;
  {\cal F}_{n,\ell}^{(0)}\, x^{\,-1} \ln^{\,n+\ell-1}x \:+\:
  {\cal F}_{n,\ell}^{(1)}\, x^{\,-1} \ln^{\,n+\ell-2}x \:+\: \dots \;\; .
\eeq
If the constants up to ${\cal F}_{n,\ell}^{(m)}$ are known for all $n$ and
$\ell$, then the splitting functions and coefficient functions can be
determined at N$^m$LL accuracy at all orders of the strong coupling.
As observed in Ref.~\cite{AV2011}, the $n^{\rm th}$ order small-$x$ 
contributions to $\widehat{F}_{T,\,\phi}$ are built up from $n$ terms of the 
form
\beq
\label{FThatDec}
  \big( A_{n,k} \,\ep^{\,-2n+1} 
  \,+\, B_{n,k} \,\ep^{\,-2n+2}
  \,+\, \ldots \big) \, x^{\,-1-2\:\!k\:\! \ep}
  \;\; , \quad k = 1,\, \dots,\, n \; .
\eeq
Since the terms with $\ep^{\,-2n+1}, \:\dots\,,\: 
\ep^{\,-n-1}$ have to cancel in sum (\ref{FThatFact}), there are $\,n\!-\!1$
relations between the LL coefficients $A_{n,k}$ which lead to the constants
${\cal F}_{n,\ell}^{(0)}$ in Eq.~(\ref{FThatLogs}), $\,n\!-\!2$ relations
between the NLL coefficients $B_{n,k}$ etc.
As discussed above, a N$^m$LO calculation fixes the (non-vanishing)
coefficients of $\ep^{\,-n}, \:\dots\,,\: \ep^{\,-n+m}$ at all orders $n$,
adding $m+1$ more relations between the coefficients in Eq.~(\ref{FThatDec}).
Consequently the highest $m\!+\!1$ double logarithms, i.e., the N$^m$LL
approximation, can be determined order by order from the N$^m$LO results.
Finally the resulting series, here calculated to order $\as^{18}$ using
{\sc Form} and {\sc TForm} \cite{FORM}, can be employed to find their 
generating functions via over-constrained systems of linear equations. 
The whole procedure is analogous to, if computationally more involved than, 
the large-$x$ resummation in Ref.~\cite{Large-x}.

\setcounter{equation}{0}
\section{$N$-space results: splitting functions $P_{\!ik}^{\,T}$}

\vspace{-2mm}
It~turns out that the resummed splitting functions in Mellin space can be 
expressed in terms of 
\bea
\label{SLdefs}
   S        \:\:=\:\: ( 1 - 4\:\!\xi )^{1/2} \quad \mbox{and} \quad
   {\cal L} \:\:=\:\: \ln \big(\, \frct{1}{2}(1+S) \big)
            \:\:=\:\: -\:\! \ln \big(\, \frct{1}{2\:\!\xi}(1-S) \big)
\eea
with $\xi = -\:\!8\,\ca\ar/\bar{N}^2$ and $\bar{N} \equiv N\!-\!1$.
At NNLL accuracy, i.e., resumming the contributions 
$\as^{\,n}\, x^{\,-1}\ln^{\,2n-a} x$ with $a = 2,\,3,\,4$ the 
(flavour-singlet) splitting functions in Eq.~(\ref{Devol}) are given by
\bea
\label{Pqq-cl}
  P_{\rm qq}^{\,T}(N) \!&\!=\!&
  \frct{4}{3}\:\*\frct{\cf\*\nf}{\ca}\:\*\ar\,\* 
  \bigg\{ \frct{1}{2\*\xi}\* (S-1)\* ({\cal L}+1) + 1 \bigg\}
\nn\\ & & \mbox{\hspn} 
  + \:\frct{1}{18}\:\*\frct{\cf\*\nf}{\cath}\:\*\ar\* \bar{N}\*\, \bigg\{ 
    ( - 11\,\*\cas + 6\,\*\ca\*\nf - 20\,\*\cf\*\nf )\:\* \frct{1}{2\*\xi}\* 
    (S-1+2\,\*\xi) + 10\,\*\cas\,\* \frct{1}{\xi}\*(S-1)\,\*{\cal L} 
\qquad \nn\\[-1mm] & & \mbox{} 
  - ( 51\,\*\cas - 6\,\*\ca\*\nf + 12\,\*\cf\*\nf )\*\:\frct{1}{2}\*(S-1) 
  + ( 11\,\*\cas + 2\,\*\ca\*\nf - 4\,\*\cf\*\nf )\*\,  S^{\,-1}\* {\cal L}
\nn\\ & & \mbox{} 
  + ( 5\,\*\cas - 2\,\*\ca\*\nf + 6\,\*\cf\*\nf )\*\,  \frct{1}{\xi}\* (S-1)\* 
    {\cal L}^2 + ( 51\,\*\cas - 14\,\*\ca\*\nf + 36\,\*\cf\*\nf )\*\, {\cal L}
    \bigg\}
\; , \\[1mm]
\label{Pqg-cl}
  P_{\rm qg}^{\,T}(N) \!&\!=\!& 
  \frct{\ca}{\cf}\, P_{\rm qq}^{\,T}(N) 
  \;-\; \frct{2}{9}\:\*\frct{\nf}{\cas}\:\*\ar\* \bar{N}\* 
  \left( \cas + \ca\*\nf - 2\,\*\cf\*\nf \right)\* 
  \bigg\{ \frct{1}{2\*\xi}\* (S-1)\* ({\cal L}+1) + 1 \bigg\}
\; , \\[1mm]
\label{Pgg-cl}
  P_{\rm gg}^{\,T}(N) \!&\!=\!& 
  \frct{1}{4}\*\, \bar{N}\* (S-1)
  \;-\; \frct{1}{6\,\*\ca}\*\, \ar\*\, ( 11\,\*\cas + 2\,\*\ca\*\nf 
  - 4\,\*\cf\*\nf )\*\, (S^{\,-1}-1)
  \;-\; P_{\rm qq}^{\,T}(N) 
\nn \\ & & \mbox{\hspn} + \:
  \frct{1}{576\*\,\cath}\*\:\ar\* \bar{N}\*\, \bigg\{
\!
   \left( [1193 - 576\,\*\z2]\* \cafo - 140\,\*\cath\*\nf + 4\,\*\cas\*\nfs 
   - 56\,\*\cas\*\cf\*\nf + 16\,\*\ca\*\cf\*\nfs 
\right. \quad \nn \\ & & \left. \mbox{} 
   - 48\,\*\cfs\*\nfs \right)\* (S-1)
   \:+\: \left( [830 - 576\*\,\z2]\*\,\cafo + 96\*\,\cath\*\nf 
   - 8\*\,\cas\*\nfs - 208\*\,\cas\*\cf\*\nf 
\right.  \nn \\ & & \left. \mbox{} 
   + 64\*\,\ca\*\cf\*\nfs - 96\*\,\cfs\*\nfs \right)\* (S^{\,-1}-1)
   \:+\: ( 11\*\,\cas + 2\*\,\ca\*\nf - 4\*\,\cf\*\nf )^2 \* (S^{\,-3}-1)
  \bigg\} 
\; , \\[1mm]
\label{Pgq-cl}
  P_{\rm gq}^{\,T}(N) \!&\!=\!& \frct{\cf}{\ca}\, P_{\rm gg}^{\,T}(N)
  \;-\; \frct{1}{3}\*\:\frct{\cf}{\cas}\*\:\ar\* \left( 
  \cas + \ca\*\nf - 2\,\*\cf\*\nf \right)\* \frct{1}{\xi}\* (S-1+2\,\*\xi)
\nn \\ & & \mbox{\hspn} + \:
  \frct{1}{36}\*\:\frct{\cf}{\cafo}\*\: \ar\* \bar{N}\* \bigg\{
\!
    \left( 11\,\*\cafo + 13\,\*\cas\*\nf\* (\ca - 2\,\*\cf) + 2\,\*\cas\*\nfs 
      - 8\*\,(\ca-\cf)\* \cf\*\nfs \right)\* (1 - S^{\,-1})
\nn \\ & & \mbox{} - \;
    \left( 48\,\*\cafo - 45\,\*\cath\*\cf - 72\,\*\z2\,\*\cath\*(\ca-\cf) 
      - 33\,\*\cath\*\nf + 2\,\*\cas\*\nfs + 48\,\*\cas\*\cf\*\nf 
\right. \nn \\ & & \left. \mbox{}
      - 8\,\*\cfs\*\nfs \right)\* \frct{1}{\xi}\* (S-1+2\,\*\xi)
  + \left( - 54\,\*\cafo + 45\,\*\cath\*\cf + 72\,\*\z2\,\*\cath\* (\ca-\cf) 
  + 23\,\*\cath\*\nf
\right. \nn \\[-1mm] & & \left. \mbox{}
      - 28\,\*\cas\*\nf\*\cf - 8\,\*(\ca-2\,\*\cf)\* \cf\*\nfs \right) \*
    \frct{1}{\xi}\* (S-1)\*{\cal L} \bigg\}
\eea
with $\cf = 4/3$ in QCD and $\z2 = \pi^2/6$.
The respective first lines of Eqs.~(\ref{Pqq-cl}) -- (\ref{Pgq-cl}) are the
LL (for $P_{\rm gq}^{\,T}$ and $P_{\rm gg}^{\,T}$, already determined in 
Ref.~\cite{LLxto0}) and NLL contributions, the rest represents the NNLL terms.
Of course, no negative powers of $\ca$ and no non-$(N\!-\!1)^{-1}$ terms remain 
when these results are expanded in powers of $\ar$. 
After combination with the LO and NLO splitting functions \cite{FPplb} 
-- with the $1/(N-1)^\ell$, $1 \leq \ell \leq 3$ poles removed to avoid double 
counting --
these results provide a combined ($N\!=\!1$ finite) all-$x$ 
`NLO $+$ resummed' evolution of the \MSb\ fragmentation distributions.
The crucial step towards deriving Eqs.~(\ref{Pqq-cl}) -- (\ref{Pgq-cl}) is 
discussed in Appendix A.

As discussed in Ref.~\cite{AV2011}, it is possible to also obtain the next
(N$^3$LL) contributions to $P_{\rm qq}^{\,T}$ and $P_{\rm qg}^{\,T}$, due to 
$A_{n,1} = B_{n,1} = \ldots = 0$ in Eq.~(\ref{FThatDec}) for 
$\widehat{F}_{\,T,\rm q}$ and $\widehat{F}_{\,\phi,\rm q}$. We have been able
to find the exact all-order expression also for these contributions. 
However, the results are considerably more lengthy than Eqs.~(\ref{Pqq-cl}) 
and (\ref{Pqg-cl}) and hence are deferred to Eqs.~(\ref{Pqq-n3}) and 
(\ref{Pqg-n3}) in Appendix B.

The first moments of the combined splitting functions receive contributions
from (\ref{Pqq-cl}) -- (\ref{Pgq-cl}), taking the limit $N \ra 1$ for fixed 
$\as$, and the `truncated' fixed-order results $\overline{P}^{\,T(m)}$ with
\beq
\label{LOmom1}
  \overline{P}^{\,T(0)}_{\rm qq} \:=\: 0 \; , \;\;
  \overline{P}^{\,T(0)}_{\rm qg} \:=\: \frct{4}{3}\:\nf \; , \;\;
  \overline{P}^{\,T(0)}_{\rm gq} \:=\: \mbox{}- 3\:\!\cf \; , \;\;
  \overline{P}^{\,T(0)}_{\rm gg} \:=\: \mbox{}
      -\:\!\frct{11}{3}\:\ca - \frct{2}{3}\:\nf
\eeq
at $N=1$. This leads to the `NLO $+$ resummed' results
\bea
\label{TOTmom1}
  P^{\,T}_{\rm qg}(N\!=\!1) \!&\!=\!&
  \frct{8}{3}\,\*\nf\,\* \ar
  \,-\, \frct{1}{3\,\*\cas}\,\* \left( 17\,\*\cas\,\*\nf - 2\,\*\ca\*\nfs 
         + 4\,\*\cf\*\nfs\, \right) \,\* (2\,\*\ca\,\* \art)^{1/2}
  \,+\, {\cal O}(\ars) \; ,
\nn \\[0.5mm]
  P^{\,T}_{\rm qq}(N\!=\!1) \!&\!=\!&
  \frct{\cf}{\ca}\:\* \Big( P^{\,T}_{\rm qg}(N\!=\!1)
  \,-\, \frct{4}{3}\,\*\nf\, \*\ar \Big) 
  \:+\: {\cal O}(\ars) \; , 
\nn \\[0.5mm]
  P^{\,T}_{\rm gg}(N=1) \!&\!=\!& 
  (2\,\*\ca\,\* \ar)^{1/2}  
  \,-\, \frct{1}{6\,\*\ca}\,\* 
    ( 11\,\*\cas + 2\,\*\ca\*\nf + 12\,\*\cf\*\nf )\,\*\ar
\nn \\[-0.5mm] & & \mbox{\hspn}
  \,+\, \frct{1}{144\,\*\cath}\,\* 
    \left( [ 1193 - 576\,\*\z2 ] \,\*\cafo - 140\,\*\cath\*\nf 
      + 4\,\*\cas\*\nfs + 760\,\*\cas\*\cf\*\nf 
\right. \nn \\ & & \mbox{} \left.
      - 80\,\*\ca\*\cf\*\nfs + 144\,\*\cfs\*\nfs\, \right)
    \,\* (2\,\*\ca\,\* \art)^{1/2}
  \,+\, {\cal O}(\ars) \; ,
\nn \\[0.5mm]
  P^{\,T}_{\rm gq}(N\!=\!1) \!&\!=\!&
  \frct{\cf}{\ca}\:\* \Big( P^{\,T}_{\rm gg}(N\!=\!1)
  \,+\, \frct{4}{3}\,\*\frct{\cf\*\nf}{\ca}\, \*\ar \Big) 
  \:+\: {\cal O}(\ars) \; .
\eea
It is interesting to note that the combination 
$\,P^{\,T}_{\rm qq}- P^{\,T}_{\rm qg}+ P^{\,T}_{\rm gq}- P^{\,T}_{\rm gg}\,$
of the resummed first moments (\ref{TOTmom1}) vanishes for $\ca=\cf$ 
irrespective of the numbers of flavours $\nf$. 
Instead inserting the QCD values for the colour factors, we obtain for $\nf= 5$
the numerical series
\bea
\label{Pres-n1}
  P_{\rm qq}^{\,T}(N\!=\!1) \!&\!\cong\!&
  \phantom{0.3071\,\as^{\,1/2} \:-\:}\;
  0.2358\, \as \:-\: 0.6773\, \as^{\,3/2} \:+\: 0.5880\, \ass \; ,
\nn \\
  P_{\rm qg}^{\,T}(N\!=\!1) \!&\!\cong\!&
  \phantom{0.3071\,\as^{\,1/2} \:-\:}\;
  1.0610\, \as \:-\: 1.5240\, \as^{\,3/2} \:+\: 1.8089\, \ass \; ,
\nn \\
  P_{\rm gq}^{\,T}(N\!=\!1) \!&\!\cong\!&
  0.3071\,\as^{1/2} \:-\: 0.3059\, \as \:+\: 0.2884\, \as^{\,3/2} \; ,
\nn \\
  P_{\rm gg}^{\,T}(N\!=\!1) \!&\!\cong\!&
  0.6910\,\as^{1/2} \:-\: 0.9240\, \as \:+\: 0.6490\, \as^{\,3/2} \; ,
\eea
with benign coefficients in terms of a rather large expansion parameter with 
$\sqrt{\as} \simeq 0.34$ at $\mu = M_Z$. In order to illustrate this behaviour
we have included the $\ass$ coefficients resulting from Eqs.~(\ref{Pqq-n3}) 
and (\ref{Pqg-n3}) together with the corresponding truncated second-order 
splitting functions $\overline{P}^{\,T(1)}_{\!\rm qi}$.

\vspace{-2mm}
\setcounter{equation}{0}
\section{Coefficient functions for $F_T$ and $F_\phi$}

\vspace{-2mm}
We now turn to the coefficient functions (\ref{Cexp}) for $F_T$ and $F_\phi$. 
Their leading and next-to-leading logarithms, 
$\as^{\,n}\, x^{\,-1}\ln^{\,2n-a} x$ with $a = 1,\,2$, 
need to be included in a `NLO $+$ resummed' approximation which is applicable 
at all values of $x$ and finite at $N=1$. The corresponding $N$-space results, 
again derived from the series expansions in Ref.~\cite{AV2011}, are given by
(using $F \,\equiv\, S^{\,-1/2\,}$)
\bea
  \label{cTg-cl}
  C_{T,\rm g}^{}(N) \!&\!=\!&
  \frct{\cf}{\ca}\,\* (F - 1) 
  \:+\:
  \frct{1}{1152}\,\*\frct{\cf}{\cath}\:\* \bar{N}\:\*
  \left\{ 
      ( - 555\,\*\cas + 66\,\*\ca\*\nf - 480\,\*\cf\,\*\nf )\* (F^{-1} - 1)
\right. \nn \\ & & \left. \mbox{}
    + ( 868\,\*\cas + 152\,\*\ca\*\nf - 336\,\*\cf\*\nf )\* (F - 1)
    + ( 11\,\*\cas - 2\,\*\ca\*\nf )\*
      \left[ 6\,\* (F^{3} - 1) 
\right. \right. \nn \\ & & \left. \left. \mbox{}
      - 5\,\* (F^7 - 1) \right]
    - ( 132\,\*\cas + 24\,\*(\ca-2\,\*\cf)\,\*\nf )\* (F^{5} - 1) 
    + 384\,\*\cf\*\nf\,\* F\* {\cal L} 
  \right\} 
\; ,
\\[0.5mm]
  \label{cTq-cl}
  C_{T,\rm q}^{}(N) \!&\!=\!& 1 \:+\: 
   \frct{1}{3}\,\* \frct{\cf\*\nf}{\cas}\,\* \bar{N}\:\* 
    ( F^{-1} -1 - F\*{\cal L} )
\eea
and
\bea
  \label{cPg-cl}
  C_{\phi,\rm g}^{}(N) \!&\!=\!& 1 \:+\: 
  \frct{\ca}{\cf}\, C_{T,\rm g}^{}(N) \:+\:
  \frct{1}{12\,\*\cas}\:\* \bar{N}\:\* 
  \left\{ ( 2\,\*\cas - \ca\*\nf + 2\,\*\cf\*\nf )\* (F^{-1} - 1)
\qquad \right. \nn \\ & & \left. \mbox{} 
   \,-\, ( 4\,\*\cas + \ca\*\nf - 2\,\*\cf\*\nf )\* (F - 1) \right\}  
 \; ,
\\[0.5mm]
  \label{cPq-cl}
  C_{\phi,\rm q}^{}(N) \!&\!=\!& 
 \frct{\ca}{\cf}\, \left( C_{T,\rm q}^{}(N) - 1 \right) 
 \; .
\eea
Here we have included, besides the resummed $1/(N\!-\!1)$ terms, the 
zeroth-order terms in Eq.~(\ref{Cexp}). The leading logarithmic contributions 
in Eqs.~(\ref{cTg-cl}) and (\ref{cPg-cl}) have been obtained before, in a 
completely different manner, in Ref.~\cite{ABKK11} 
(where, as in Refs.~\cite{RvN-FTL,MMoch06}, the normalization of $C_{a,\rm g}$
in SIA differs by a factors of two from that adopted in the present article
and Ref.~\cite{AMV1}).
We have also been able to derive closed $N$-space resummations of the NNLL 
contributions, $\as^{\,n}\, x^{\,-1}\ln^{\,2n-3} x$, to all four quantities. 
The corresponding more lengthy expressions can be found in Appendix B.

The first moments of Eqs.~(\ref{cTg-cl}) -- (\ref{cPq-cl}) vanish except for
$\as^{\,0}$ contributions in Eqs.~(\ref{cTg-cl}) and (\ref{cTq-cl}). Note, in 
particular, that the zeroth-order contribution (\ref{Cexp}) is compensated in 
$C_{\phi,\rm g}^{}$ by the resummation of the leading logarithms.
The first $\as$-dependent contributions are due to Eqs.~(\ref{cTq-nn}) -- 
(\ref{cPhiq-nn}) and the truncated NLO coefficient functions 
$\overline{c}_{a,i}^{\,(1)}$ (i.e., the corresponding quantities of 
Eq.~(\ref{Cexp}) with the $1/(N\!-\!1)$ poles removed) and read
\bea
  \label{cTqg-N1}
  C_{T,\rm q}^{}(N\!=\!1) \!&\!=\!& 
  -\,\frct{\ca}{\cf}\, C_{T,\rm g}^{}(N\!=\!1) \,+\: \ldots 
\nn \\ \!&\!=\!& 
  1 \:+\: \frct{1}{9}\,\*\frct{\cf}{\cath}\* 
    \left( 9\,\*\cath \,+\, 50\,\*\cas\*\nf 
    \,-\, 8\,\*(\ca - 2\,\*\cf)\* \nfs\, \right)\* \ar 
  +\, \ldots \; , 
\\[-1mm]
  \label{cPqg-N1}
  C_{\phi,\rm q}^{}(N\!=\!1) \!&\!=\!&
  -\,\frct{\ca}{\cf}\, C_{\phi,\rm g}^{}(N\!=\!1) \,+\: \ldots \:
  \:=\;
  -\,\frct{4}{9}\,\*\frct{\nfs}{\cas}\* ( \ca - 2\,\*\cf )\,\* \ar
  +\, \ldots \:\; .
\eea
The next corrections at $N=1$ may be expected at order $\ass$ which is beyond 
our present reach.


All our results above are given in the standard \MSb\ factorization scheme.
The transformation to another scheme $S$ is given by
\beq
\label{Strf}
  C_S(N) \:\: = \:\: C(N)\, [Z_S(N)]^{\,-1} \;\; ,\quad
  P_S(N) \:\: = \:\: \Big( Z_S(N)\, P(N) 
   + \beta\, \frct{\partial Z_S(N)}{\partial \ar}\,\Big) [Z_S(N)]^{\,-1} \; ,
\eeq
where (suppressing the Mellin variable and the scheme index) $C$ and $P$ 
represent the matrices 
\beq
\label{CPmat}
  C \;=\; \left( \begin{array}{cc}
          \!  C_{T,\rm q}^{}\!    & C_{T,\,\rm g}^{} \! \\[1mm]
          \!  C_{\phi,\rm q}^{}\! & C_{\phi,\rm g}^{} \!
          \end{array} \right)  
  \;\; \mbox{ and } \;\;
  P \;=\;\: \left( \begin{array}{cc}
         P_{\rm qq}^{\,T} & P_{\rm gq}^{\,T} \\[2mm]
         P_{\rm qg}^{\,T} & P_{\rm gg}^{\,T}
   \end{array} \right)
  \;\; ,\; \mbox{ with } \;\;
  P_{\,\rm LL}^{} \;=\;\: \left( \begin{array}{cc} 
         0 & \frct{\cf}{\ca}\,p_{\:\!\rm LL}^{} \\[1mm]
         0 & p_{\:\!\rm LL}^{}
  \end{array} \right)  
\eeq
in \MSb. Eq.~(\ref{Strf}) includes, for $Z_S = C$, the transformation to the
`physical evolution kernels' $K$,
\beq
\label{KTphi}
  \frac{\partial}{\partial\ln \Qs}
  \left( \begin{array}{c} \!F_T\! \\ \!F_\phi\! \end{array} \right)
  =
  \left( \begin{array}{cc}
         \! K_{TT}     \! & K_{T\phi} \! \\
         \! K_{\phi T} \! & K_{\phi\phi}\! 
   \end{array} \right)
  \left( \begin{array}{c} \!F_T\! \\ \!F_\phi\! \end{array} \right) \; ,
\eeq
for the system $F = (F_T,\,F_\phi)$ of fragmentation functions [for $N\neq 1$, 
since ${\rm det}\; C$ vanishes for the results in Eqs.~(\ref{cTqg-N1}) and 
(\ref{cPqg-N1})], cf.~Ref.~\cite{FP82}.
At LL accuracy this transformation, as well as that to the massive-gluon
scheme of Ref.~\cite{Mueller83}, see also Ref.~\cite{ABKK11}, is of the form
\beq
\label{ZsLL}
  Z_{S,\rm LL}^{} \:\:=\:\: \frac{1}{1 + c_{S,\rm LL}^{}}
   \left( \begin{array}{cc} \!
    1+c_{S,\rm LL}^{}\;& \frct{\cf}{\ca}(c_{\rm LL}^{}-c_{S,\rm LL\,}^{})\! 
    \\[2mm] 0          & 1+c_{\,\rm LL}^{}
   \end{array} \right) 
\eeq
with $c_{\rm LL}^{} \,=\, F-1$ [recall $F = S^{\,-1/2}$ with $S$ given in 
Eq.~(\ref{SLdefs})], $c_{S,\rm LL}^{} \,=\, \frac{1}{2}\,(S^{\,-1}-1)$ for 
the scheme of Ref.~\cite{Mueller83} and $c_{S,\rm LL}^{} \,=\, 0$ for the 
physical kernels. $P_{\,\rm LL}^{}$ is invariant for this form of $Z_{S\,}$. 
Since we do not know the coefficient functions in the massive gluon scheme 
beyond LL accuracy, we are not able to use the scheme transformation to 
compare our results to the NLL splitting functions in Ref.~\cite{Mueller83}.
The transformation to the physical kernels in Eq.~(\ref{KTphi}), on the other 
hand, does not pose problems. 
It is interesting to note that, disregarding overall prefactors, the results
to NNLL (for $K_{\phi T}$ and $K_{\phi\phi}$) and N$^3$LL (for $K_{TT}$ and 
$K_{T\phi}$) accuracy include only integer coefficients in the expansion in 
powers of $\as$. 
Consequently terms with the logarithm ${\cal L}$ in Eq.~(\ref{SLdefs}) do not
enter their closed all-order expressions, in contrast to Eqs.~(\ref{Pqq-cl}) 
-- (\ref{Pgq-cl}) for the \MSb\ splitting functions. 

\vspace{-2mm}
\setcounter{equation}{0}
\section{Coefficient functions for $F_L$}

\vspace{-2mm}
The remaining SIA coefficient functions with $x^{\,-1}\ln^{\,k}x$ 
contributions are those for the longitudinal fragmentation function $F_L$. 
There is no $\as^{\,0}$ term in this case, hence the $\as^{\,n}$ results are 
required for the N$^{n\!-\!1}$LO approximation and thus for resummation of the
$n$ highest logarithms outlined above.
Since the NLO coefficient function $C_{L,\rm g}$ includes terms up to
$x^{\,-1}\ln^{\,2} x$, the NNLL resummation is required to extend the all-$x$ 
`NLO + resummed' approximation to $F_L$.
The coefficient functions for $F_L$ have been calculated so far only to order 
$\ass$ \cite{RvN-FTL,MMoch06}. It is, however, possible to derive (at least) 
their highest three $\as^{\,3}\, x^{\,-1}\ln^{\,k} x$ contributions using 
$x \ra 1/x$ analytic-continuation (${\cal AC}$) relations between DIS and SIA 
physical kernels along the lines of Ref.~\cite{AMV1}, see also 
Ref.~\cite{BRvN00}.

For this purpose we have calculated the evolution kernels $K^{T}$ and $K^{S}$
corresponding to Eq.~(\ref{KTphi}) for the system $(F_T,\,{\widetilde F}_L)$ 
with 
$\,{\widetilde F}_L(N,\Qs) \,=\, F_L(N,\Qs)/(\as\,c_{L,\rm q}^{\,(1)}(N))\,$
and its `spacelike' counterpart of DIS structure functions 
$(F_1^{},\,{\widetilde F}_L^{\,S})$ for which all NNLO coefficient functions 
are known \cite{ZvN-F2L,MVerm99,MVV6}. At this order these kernels are expected
to be related by
\beq
\label{AC-KTL}
   K_{ab}^T(x,\as) \:\:=\:\: {\cal AC} \left[K_{ab}^S(x,\as)\right] 
\eeq
up to terms beyond the presently required logarithmic accuracy.
These relations form a check of the respective two highest logarithms in 
$c_{L,\rm q}^{\,(3)}$ and $c_{L,\rm g}^{\,(3)}$, which have been derived 
by resumming the NLO results in Ref.~\cite{AV2011}, and provide an 
over-constrained system of four equations for the hitherto unknown 
coefficients of $\,\as^{\,3}\, x^{\,-1}\ln^{\,2} x$ in $C_{L,\rm g}\,$ and
$\,\as^{\,3}\, x^{\,-1}\ln x$ in $C_{L,\rm q}\,$. Eq.~(\ref{AC-KTL}) leads to
\bea
\label{cLq3}
  c_{L,\rm q}^{\,(3)}(N) &\!=\!&
    256\,\*\ca\*\cf\*\nf\*\: \bar{N}^{-4} \:-\:
    64\,\*\cf\* \left( \frct{17}{3}\,\*\ca\*\nf - \frct{2}{3}\,\*\cf\*\nf 
    -\frct{4}{9}\,\*\nfs \right)\* \bar{N}^{-3}
\nn \\ & & \mbox{} - \:
   64\,\*\cf\* \left( \frct{94}{27}\,\*\ca\*\nf 
   - \left[\frct{5}{6} - \frct{4}{3}\,\*\z2 \right]\* \cf\*\nf
   - \frct{1}{3}\,\*\nfs \right)\* \bar{N}^{-2} 
   \:+\: {\cal O}(\bar{N}^{-1}) \; ,
\\[1mm]
\label{cLg3}
  c_{L,\rm g}^{\,(3)}(N) &\!=\!& \mbox{}
    1408\,\*\cas\*\cf \*\: \bar{N}^{-5} \:-\:
    32\,\*\cf\* \left( \frct{241}{3}\,\*\cas - 10\,\*\ca\*\cf 
    - \frct{20}{3}\,\*\cf\*\nf - \frct{2}{3}\* \ca\*\nf
    \right)\* \bar{N}^{-4} 
\quad \nn \\ & & \mbox{} 
    + 32\,\*\cf\* \left( \left[ 19 - 22\,\*\z2 \right]\*\cas 
    - \frct{17}{6}\* \ca\*\cf + \frct{61}{18}\,\*\ca\*\nf
    - \frct{37}{9}\* \cf\*\nf \right)\* \bar{N}^{-3}
   \:+\: {\cal O}(\bar{N}^{-2}) \; . \quad 
\eea
The resulting NNLL resummation for $F_L$ -- which does not involve 
Eq.~(\ref{cLq3}) -- can be written as
\bea
\label{cLq-cl}
  C_{L,\,\rm q}^{}(N) \!&\!=\!&
  \:\frct{2}{3}\:\*\frct{\cf\*\nf}{\ca}\:\*\ar\,\*
  \left\{\frct{1}{\xi}\*\, (S-1)\*( F\*{\cal L} -F^{-1} )
  - 2 \right\}
  \nn\\ & & \mbox{\hspn} 
  + \frct{1}{1728}\:\*\frct{\cf\*\nf}{\cath}\:\*\ar\*\bar{N}\,\* 
  \Big\{
    96\* \left( 5\,\*\cas - 2\,\*\ca\*\nf + 6\,\*\cf\*\nf \right)\*
    \frct{1}{\xi}\*(S-1)\*F\*{\cal L}^2 
  - \left( 231\*\cas  - 42\,\*\ca\*\nf 
  \right. \nn \\[-1mm] & & \left. \mbox{}
  - 32\,\*\ca\*\cf + 128\,\*\cf\*\nf\right)\,\* 9\,\*
    \frct{1}{\xi}\*(S-1)\*\,F^{-1}\*{\cal L} 
  + 8\,\*\left(53\,\*\cas + 28\,\*\ca\*\nf + 36\,\*\ca\*\cf 
  \right. \nn \\[-1mm] & & \left. \mbox{}
  - 114\,\*\cf\*\nf\right)\*\frct{1}{\xi}\*(S-1)\*\,F\*{\cal L} 
  - 6\,\*\left(55\,\*\cas + 6\,\*\ca\*\nf - 16\,\*\cf\*\nf\right)
      \*\frct{1}{\xi}\*(S-1)\*\,F^{3}\*{\cal L}
  \nn \\[-1mm] & & \mbox{}
  - 24\*\left(11\,\*\cas - 2\,\*\cf\*\nf \right)\*
     \frct{1}{\xi}\*(S-1)\*\,F^{5}\*{\cal L}  
  - 5\*\left(11\,\*\cas - 2\,\*\ca\*\nf \right)\*
     \left(\frct{1}{\xi}\*(S-1)\*\,F^{7}\*{\cal L} 
  \right. \nn \\[-1mm] & & \left. \mbox{}
    + 4\*\,(F^{5}-1)\right) 
   - 16\*\left( 71\,\*\cas + 10\,\*\ca\*\nf - 48\,\*\cf\*\nf \right)\*
      \frct{1}{\xi}\*\,\left( (S-1)\*F^{-1} + 2\,\*\xi\, \right)
  \nn \\[-1mm] & & \mbox{}
  - 12\*\left( 577\,\*\cas - 54\,\*\ca\*\nf - 96\,\*\cf\*\ca 
      + 144\,\*\cf\*\nf \right)\*(F^{-1}-1)
  + 4\*\left( 755\,\*\cas 
  \right. \nn \\[-1mm] & & \left. \mbox{}
  - 146\,\*\ca\*\nf + 384\,\*\cf\*\nf\right)\*(F-1) 
  - 4\*\left( 209\,\*\cas + 10\,\*\ca\*\nf 
    - 48\,\*\cf\*\nf\right)\*(F^{3}-1) 
  \Big\}
\; , \\[1mm]
\label{cLg-cl}
  C_{L,\rm g}^{}(N) \!&\!=\!& 
  \frct{1}{4}\,\*\,\frct{\cf}{\ca}\*\bar{N}\*(F^{-1}-F) 
  \:-\:\frct{\cf}{\ca}\: c_{L\,\rm q}^{\,T}(N)
  \nn \\[0.5mm] & & \mbox{\hspn}
  + \frct{1}{144}\,\*\frct{\cf}{\cas}\,\*\ar\*
  \Big\{
  16\*\left( 29\,\*\cas + \ca\*\nf - 3\,\*\cf\*\nf \right)\*
      ( \frct{1}{\xi}\,\* (S-1)\*F^{-1} + 2 )
  + \left( 737\,\*\cas + 58\,\*\ca\*\nf
 \right. \nn \\[-0.5mm] & & \left. \mbox{}
    + 288\,\*\ca\*\cf - 192\,\*\cf\*\nf \right)\*(F-1)
  - \left( 121\,\*\cas + 26\,\*\ca\*\nf - 48\,\*\cf\*\nf \right)\*(F^{3}-1)
  \nn \\[-0.5mm] & & \mbox{}
  - \left( 209\,\*\cas + 10\,\*\ca\*\nf - 48\,\*\cf\*\nf \right)\*(F^{5}-1)
  - 5\*\left( 11\,\*\cas - 2\,\*\ca\*\nf \right)\*(F^{7}-1)
  \Big\}
  \nn \\ & & \mbox{\hspn}
  +\frct{1}{331776}\,\*\frct{\cf}{\cafo}\,\*\ar\*\bar{N}\,\*\Big\{
   - 256\* \left( [104 - 4212\,\*\z2]\,\*\cafo - 459\,\*\cath\*\cf 
     + 533\,\*\cath\*\nf - 480\,\*\cas\*\cf\*\nf 
 \right. \nn \\[-0.5mm] & & \left. \mbox{}
     - 16\,\*\cas\*\nfs - 60\,\*\cf\*\nfs\,\*[\ca -3\,\*\cf] \right)\*
     (\frct{1}{\xi}\*(S-1)\*F^{-1}+2)
  - 36864\,\*\cf\*\nf\*(\cas + \ca\*\nf 
  \nn \\[-1mm] & & \mbox{}
    - 2\,\*\cf\*\nf )\,\*(\frct{1}{\xi}\*(S-1)\*F\*{\cal L})
  + \left( [ 221217 + 101376\,\*\z2 ]\,\*\cafo 
     - 77760\,\*\cath\*\cf - 3852\,\*\cath\*\nf 
 \right. \nn \\[-0.5mm] & & \left. \mbox{}
     - 9856\,\*\cas\*\cf\*\nf 
     - 512\,\* [12\,\*\ca\*\cfs\*\nf + \ca\*\cf\*\nfs - 3\,\*\cfs\*\nfs ]
     - 92\,\*\cas\*\nfs \right)\:\*9 \,\*(F^{-1}-1)
  \nn \\[0.5mm] & & \mbox{}
   - \left( [ 100985 - 1575936\,\*\z2 ]\,\*\cafo
       - 451584\,\*\cath\*\cf + 234740\,\*\cath\*\nf 
       - 160032\,\*\cas\*\cf\*\nf 
 \right. \nn \\[0.5mm] & & \left. \mbox{}
       + 193536\,\*\ca\*\cfs\*\nf - 14524\,\*\cas\*\nfs 
       + 87744\,\*\ca\*\cf\*\nfs - 124416\,\*\cfs\*\nfs \right) \*(F-1)
  \nn \\[0.5mm] & & \mbox{}
  - 3\* \left( 116329\,\*\cafo + 55296\,\*\z2\,\*\cafo + 23668\,\*\cath\*\nf 
    + 1924\,\*\cas\*\nfs - 12672\,\*\cath\*\cf
 \right. \nn \\[0.5mm] & & \left. \mbox{}
    - 55776\,\*\cas\*\cf\*\nf - 12480\,\*\ca\*\cf\*\nfs + 16128\,\*\cfs\*\nfs 
      \right) \*(F^{3}-1)
  - 3\* \left( 126471\,\*\cafo
 \right. \nn \\[0.5mm] & & \left. \mbox{}
    + 55296\,\*\z2\,\*\cafo + 380\,\*\cas\*\nf\*(5\,\*\ca + \nf) 
    + 25344\,\*\cath\*\cf - 50752\,\*\cas\*\cf\*\nf 
 \right. \nn \\[0.5mm] & & \left. \mbox{}
    - 9344\,\*\ca\*\cf\*\nfs - 9216\,\*\ca\*\cfs\*\nf + 17664\,\*\cfs\*\nfs 
    \right)\*(F^{5}-1)
  - 3\* \left( 71511\,\*\cafo 
 \right. \nn \\[0.5mm] & & \left. \mbox{}
    - 18260\,\*\cath\*\nf - 1348\,\*\cas\*\nfs + 10560\,\*\cath\*\cf 
    - 9664\,\*\cas\*\cf\*\nf + 6016\,\*\ca\*\cf\*\nfs 
  \right. \nn \\[0.5mm] & & \left. \mbox{}
    - 2304\,\*\cfs\*\nfs \right)\*(F^{7}-1)
  + \left( 123541\,\*\cafo + 55396\,\*\cath\*\nf - 2636\,\*\cas\*\nfs 
    - (10\,\*\ca\*\cf\*\nfs
  \right. \!\quad \nn \\[0.5mm] & & \left. \mbox{}
    +209\,\*\cas\*\cf\*\nf)\,\*480 + 11520\,\*\cfs\*\nfs \right)\*(F^{9}-1)
  + \left( 4477\,\*\cafo - 572\,\*\cath\*\nf - 44\,\*\cas\*\nfs 
  \right. \nn \\[-0.5mm] & & \left. \mbox{}
    - 96\,\*\ca\*\cf\*\nf\* (11\,\*\ca - 2\,\*\nf) \right)\*\:35\,\*(F^{11}-1)
  + 385\,\*\cas\*( 11\,\*\ca - 2\,\*\nf)^2 \*(F^{13}-1)
  \Big\}
  \; .
\eea
Eqs.~(\ref{cLq-cl}) and (\ref{cLg-cl}) complete the results required for the 
`NLO + resummed' all-$x$ flavour-singlet evolution of the fragmentation 
distributions (\ref{Devol}) and fragmentation functions (\ref{d2sigma}).
It is interesting to note that, unlike the corresponding splitting functions
and kernels for the system $(F_T,\,F_\phi)$, the upper-row elements $K_{Ta}$ 
of the resummed evolution matrix for $(F_T,\,\widetilde{F}_L)$ are suppressed 
by two powers of $\bar{N}=N\!-\!1$ with respect to their lower-row counterparts 
$K_{La}$. Also these kernels are found to include only integer coefficients 
after expansion in powers of $\ar = \as/(4\pi)$ or $\,\as/\pi$.

\setcounter{equation}{0}
\section{DMS relation and `non-singlet' resummation of $P_{\!\rm gg}^{\,T}$}

\vspace{-2mm}
Another connection between the final-state (`timelike') and initial-state
(`spacelike') evolution has been suggested in Ref.~\cite{DMS05}.
The evolution of the flavour non-singlet fragmentation distributions and
parton distributions, respectively denoted as $f_{1}^{}$ and $f_{-1}^{}$, 
is written as 
\beq
\label{DMSevol}
  \frct{\partial}{\partial \ln \Qs} \, f_\sigma(x,\Qs) \:\: = \:\: \left[ 
  P_u \!\left( \as (\Qs) \right) \otimes f_\sigma ( z^{\,\sigma}\Qs ) 
  \right](x)
\eeq
(the convolution $\otimes$ has been defined in Eq.~(\ref{Mconv}) above),
and the modified splitting functions $P_u$ are postulated to be identical for 
the timelike and spacelike cases. Eq.~(\ref{DMSevol}) has been applied and 
verified for the NNLO non-singlet \cite{MMV06} and gluon-gluon \cite{MV2} 
splitting functions. 
In the latter case its applicability is obvious for Quantum Gluodynamics, 
$\nf=0$, but it is found to hold for all terms in the limit $\cf=0$. 
In $N$-space Eq.~(\ref{DMSevol}) leads~to \cite{DMS05}
\beq
  \partial_{\,\ln \Qs}\, f_{\sigma}(N,\Qs) 
  \:\: = \:\: P_\sigma(N)\, f_{\sigma}(N,\Qs) 
  \:\: = \:\: P_u (N + \sigma\: \partial_{\,\ln \Qs})\, f_{\sigma}(N,\Qs) \; .
\eeq
The solution of this equation, obtained by expanding in powers of $\sigma$, 
can be cast in the (new) form 
\beq
  P_\sigma(N) \:\:=\:\: P_u(N) + \sum_{n=1}^{\infty}\: 
  \frac{\sigma^{\,n}}{n!}\: \frac{\partial_{}^{\,n-1}}{\partial N_{}^{n-1}} 
  \left( \frac{\partial P_u}{\partial N}\, [P_u(N)]^n\! \right)
\eeq
suitable for all-order considerations. The above relations imply that the 
difference $\delta\:\!P \,=\, P_{\,1} - P_{-1}$ of the time- and spacelike 
splitting functions at any order is given by lower-order quantities.

The crucial point in the present context is that the spacelike splitting
functions are only single-logarithmically enhanced with the 
(scheme-independent) leading terms \cite{LL-BFKL1,LL-BFKL2}
\beq
\label{PSggLL}
  P^{\,S}_{\rm gg}(N) \:\:=\:\: \frct{\ca\as}{\pi}\: \bar{N}^{\,-1}
  \,+\, 2\:\!\z3 \left( \frct{\ca\as}{\pi} \right)^{\!4} \bar{N}^{\,-4}
  \,+\, {\cal O} \:\!( \as^{\,6}\,\bar{N}^{\,-6\,} ) \; .
\eeq
Therefore the above differences and the timelike splitting functions are 
equal, $\,\delta\:\!P_{\rm gg}^{\,(n)} \,=\, P_{\rm gg}^{\,T(n)}\!$, for  
$\,n \,\geq\, \ell +1\,$ at the level of the N$^{\:\!\ell\:\!}$LL small-$x$ 
double logarithms.
The LL, NLL and NNLL contributions to $P_{\rm gg}^{\,T(n)}\!$ for $\cf = 0$ 
are thus fixed by Eq.~(\ref{DMSevol}) to all orders in $\as$ without any 
spacelike input beyond the NNLO splitting functions \cite{MVV34}, and are 
found to be identical to the results presented in Eq.~(\ref{Pgg-cl}).
As already discussed in Ref.~\cite{March06},
the N$^3$LL terms are fixed if Eq.~(\ref{PSggLL}) is used in addition, and
finally the N$^4$LL corrections require only one additional coefficient, that 
of $\as^{\,4}\, \bar{N}^{\,-3}$, of $P^{\,S}_{\rm gg}(N)$. 
However, this coefficient is presently not available in the literature for the
\MSb\ scheme.

The resulting prediction for the N$^3$LL contribution to $P_{\rm gg}^{\,T}$ 
reads
\bea
\label{Pgg-n3}
  \lefteqn{ \left. P_{\,\rm gg}^{\,T}(N) \right|^{\,\cf=0}_{\,\rm N^3LL} 
  \:\:=\:\:
  \frct{1}{13824}\,\*\frct{1}{\ca}\,\*\ar^2\*
  \Big\{4\,\* ( [\, 48125 - 3168\*\,\z2 - 15552\*\,\z3 ]\,\*\cath 
    - [\,46 + 2880\*\,\z2]\,\*\cas\*\nf
   } \nn\\&& \mbox{}
   + 236\,\*\ca\*\nfs - 8\,\*\nft )\,\*\frct{1}{\xi}\*(S-1+2\,\*\xi)
  - ([\,48473 + 9504\*\,\z2 - 25920\*\,\z3]\,\*\cath - 340\,\*\ca\*\nfs 
  \nn\\&&  \mbox{}
     + 24\,\*\nft - [\,2670 - 4032\*\,\z2]\,\*\cas\*\nf )\,\*
     \frct{1}{\xi^2}\*(S-1 +2\,\*\xi+2\,\*\xi^2)
  + 2\*([\,20337 + 7920\*\,\z2 
  \nn\\&& \mbox{}
    - 12960\*\,\z3]\,\*\cath
  - [\,2330 - 2592\*\,\z2]\,\*\cas\*\nf - 156\,\*\ca\*\nfs 
    + 24\,\*\nft)\*\frct{1}{\xi^2}\*(S^{\,-1}-1-2\,\*\xi-6\,\*\xi^2)
  \nn\\&& \mbox{}
  + 4\*([\,1617 - 1584\*\,\z2]\,\*\cath
  + [316 - 288\*\,\z2]\,\*\cas\*\nf - 40\,\*\ca\*\nfs - 8\,\*\nft)\*
    \frct{1}{\xi^2}\*(S^{\,-3}-1
  \nn\\[-1mm]&& \mbox{}
    -6\,\*\xi-30\,\*\xi^2)
  + (11\,\*\ca + 2\,\*\nf)^3\*\frct{1}{\xi^2}\*
    (S^{\,-5}-1-10\,\*\xi-70\,\*\xi^2)
  \bigg\}
 \; .
\eea
Combined with the first moment of the truncated NLO splitting function,
\beq
\label{Pgg1N1}
  \overline{P}_{\,\rm gg}^{\,T(1)}(N\!=\!1) \:\:=\:\:
  \frct{160}{27}\,\*\cas 
  \,+\, \left ( \frct{76}{27} + \frct{16}{3}\,\*\z2\! \right) \*\ca\*\nf 
  \,-\,\frct{88}{3}\,\*\cf\*\nf \; ,
\qquad
\eeq
this results yields
\beq
\label{Pgg-n3-N1} \left.
 P_{\,\rm gg}^{\,T}(N\!=\!1)\right|_{\ars}^{\cf = 0} \;=\:\: \mbox{}
   - \left( \frct{35}{3} - \frct{11}{3}\,\*\z2 - 6\,\*\z3
\! 
   \right) \* \cas
   \,-\,\left( \frct{11}{9} - \frct{10}{3}\,\*\z2
\!
    \right) \*\ca\*\nf 
\; .
\eeq
Including also the N$^4$LL contribution in Eq.~(\ref{Pgg-n4-N1}), the 
numerical result for $\nf = 5$ is given by
\bea
\label{Pggres-n1}
  \left. P_{\,\rm gg}^{\,T}(N\!=\!1) \right|^{\,\cf = 0} \!&\!\cong\!&
    0.6910\,\as^{1/2} \:-\: 0.5703\,\as \:+\: 0.0267\,\as^{3/2} 
    \:+\: 0.4946\, \ass    
\nn \\ & & \mbox{\hspn} \;\;\;
    \:+\: \Big( 1.0036 + 0.15753 \left. B_{\rm gg}^{\,S\,(3)} 
    \right|_{\,\rm NLL}^{\,\cf = 0\,} \Big) \,\as^{5/2} 
    \:+\: {\cal O} (\asth)
  \; ,
\eea
where $B_{\rm gg}^{\,S\,(3)}|_{\,\rm NLL\,}$ is the above-mentioned 
next-to-leading small-$x$ coefficient at order $\asfo$, cf.\ Eq.\ (\ref{Pexp}), 
in the notation of Eq.~(\ref{PSggLL}), i.e., the coefficient of 
$( \ca\,\as/\pi )^4\: \bar{N}^{\,-3}$.
One may expect it to be negative and no larger than a few times the 
corresponding LL coefficient $2\,\z3 \simeq 2.4041$.
An explicit determination of this coefficient and its fifth-order counterpart 
from the results of Ref.~\cite{NL-BFKL} should definitely be possible, in
particular in view of the calculations performed in 
Refs.~\cite{NL-Trf1,NL-Trf2}. 
On top of these quantities, and the known NNLO splitting function \cite{MV2} 
and sixth-order coefficient in Eq.~(\ref{PSggLL}), the $\asth$ coefficient 
in Eq.~(\ref{Pggres-n1}) requires the NNLL fourth-order term in 
Eq.~(\ref{PSggLL}), as already discussed in Ref.~\cite{March06}, which 
currently appears to be out of reach. 

\vspace{-2mm}
\setcounter{equation}{0}
\section{Analytic and numerical results in $x$-space}

\vspace*{-2mm}
We finally return to the now complete `NLO$\:+\:$resummed' approximations for 
the \MSb\ splitting functions and coefficient functions in Eqs.~(\ref{Devol}) 
and (\ref{Fconv}). Their numerical consequences in \mbox{$x$-space} can be 
obtained either by expanding the above $N$-space results to any desired order 
in $\as$ and using 
\beq
\label{Mlogs}
  \int_0^1\! dx \: x^{\,N-2}\, \ln^{\,k}x \:\:=\:\:
  (-1)^k\, \frac{k!}{(N\!-\!1)^{k+1}} \:\: ,
\eeq
or by subjecting the $N$-space expression to a standard numerical Mellin 
inversion, cf, e.g., Ref.~\cite{Pegasus}. The former can also be used to 
identify the closed form of $x$-space results such as
\bea
\label{dPgg-x} \lefteqn{ \left.
       x\:\!P_{\!\rm gg}^{\,T}(x,\as) 
 \,+\, x\:\!P_{\!\rm qq}^{\,T}(x,\as) \right|_{\,\rm NNLL} \;=\; 
  \left\{ 4\,\ca\,\ar \:+\: \frct{8}{3}\, 
   ( 11\,\cas + 2\,\ca\nf - 4\,\cf\nf)\, \ars \ln \frct{1}{x} 
  \right\}\, \frct{2}{z}\, J_1(z) }
\nn \\[1mm] & & \; \mbox{}
  + \left\{ 
    \frct{4}{9}\, ( 26\,\cf\nf - 23\,\ca\nf)\, \ars
    + \frct{8}{9\,\ca}\, ( 11\,\cas + 2\,\ca\nf - 4\,\cf\nf)^2 
    \art \ln^{\,2\!} \frct{1}{x} \right\}\, \frct{2}{z}\, J_1(z) 
\\[1mm] & & \; \mbox{}
  + \frct{32}{9\,\ca} \left( [ 134 - 72\,\z2] \,\cafo + 23\,\cath\nf
  - 48\,\cas\cf\nf + 4\,\ca\cf\nfs - 8\,\cfs\nfs \right) 
  \art \ln^{\,2\!} \frct{1}{x}\; \frct{4}{z^2}\, J_2(z)
\;\; \nn 
\eea
and 
\bea
\label{dPgq-x}
  \left.  x\:\! P_{\,\rm gq}^{\,T}(N) 
  \,-\, \frct{\cf}{\ca}\: x\:\!P_{\!\rm gg}^{\,T}(x,\as) \right|_{\,\rm NLL}
  \;=\; \mbox{} -
  \frct{32}{3}\:\frct{\cf}{\ca}\, (\cas+\ca\nf-2\,\cf\nf)\,\ars \ln \frct{1}{x} 
    \; \frct{4}{z^2}\, J_2(z) 
\eea
with
\beq
\label{zdef}
   z \;=\; ( 32\,\ca\,\ar )^{1/2}\, \ln \frct{1}{x}
\eeq
in terms of the Bessel function of the first kind $J_1(z)$ and $J_2(z)$, 
cf.~Ref.~\cite{AbrStg}.
The first line of Eq.~(\ref{dPgg-x}) represents the LL and NLL terms, and the 
next two lines the NNLL contribution. The missing ingredients for a completely 
analytical Mellin (Laplace) inversion of Eqs.~(\ref{Pqq-cl}) -- (\ref{Pgq-cl}) 
are the inverses of products of powers of the square root $S$ and the logarithm
${\cal L}$ defined in Eq.~(\ref{SLdefs}).
Both this logarithm, which also corresponds to a Bessel function in $x$-space
(see Eq.~(\ref{IMLog}) below), and the fourth root $F$ in Eq.~(\ref{cTg-cl}),
which corresponds to a hypergeometric function \cite{ABKK11}, can be viewed as
`artifacts' of the \MSb\ scheme, as neither remains after transformation to
the matrix of physical evolution kernels (\ref{KTphi}) for the fragmentation 
functions $F_T$ and $F_\phi$ and their counterparts for the system 
$(F_T,\,{\widetilde F}_L)$ discussed above Eq.~(\ref{AC-KTL}).

The result (\ref{dPgg-x}) for $P_{\rm gg}^{\,T}$ indicates a general 
single-logarithmic enhancement of the Bessel-function oscillations at 
extremely small $x$, where furthermore the contribution of $J_2(z)$ is 
suppressed by a factor $2/z$ with, e.g., 
$\,z \,\simeq\, 0.96\, \ln \frct{1}{x}\,$ for $\,\as \,\simeq 0.12$ 
-- recall that $J_1(z)$ and $J_2(z)$ differ for large $z$ only by a phase 
shift of $\pi/2$ \cite{AbrStg}.  
The asymptotically dominant 
$\ar (\ar \ln \frct{1}{x})^\ell\:\frac{2}{z}\,J_1(z)$
terms arise from the $S^{\,-2\,\ell+1}$ contributions to the N$^\ell$LL terms
and have closely related prefactors proportional to 
$(11\,\cas - 2\,\ca\nf + 4\,\cf\nf)^\ell$, see Eqs.~(\ref{Pgg-cl}), 
(\ref{Pgg-n3}) and (\ref{Pgg-n4}), which seem to point towards the possibility 
of a `second resummation'.
Taking into account the asymptotic $z^{\,-1/2}$ suppression of $J_n(z)$, the
oscillation amplitudes of these contributions to the LL, NLL and NNLL results
behave as $(\ln \frct{1}{x})^{-3/2}$, $(\ln \frct{1}{x})^{-1/2}$ and 
$(\ln \frct{1}{x})^{1/2}$, respectively. 

\begin{figure}[p]
\vspace*{-1mm}
\centerline{\epsfig{file=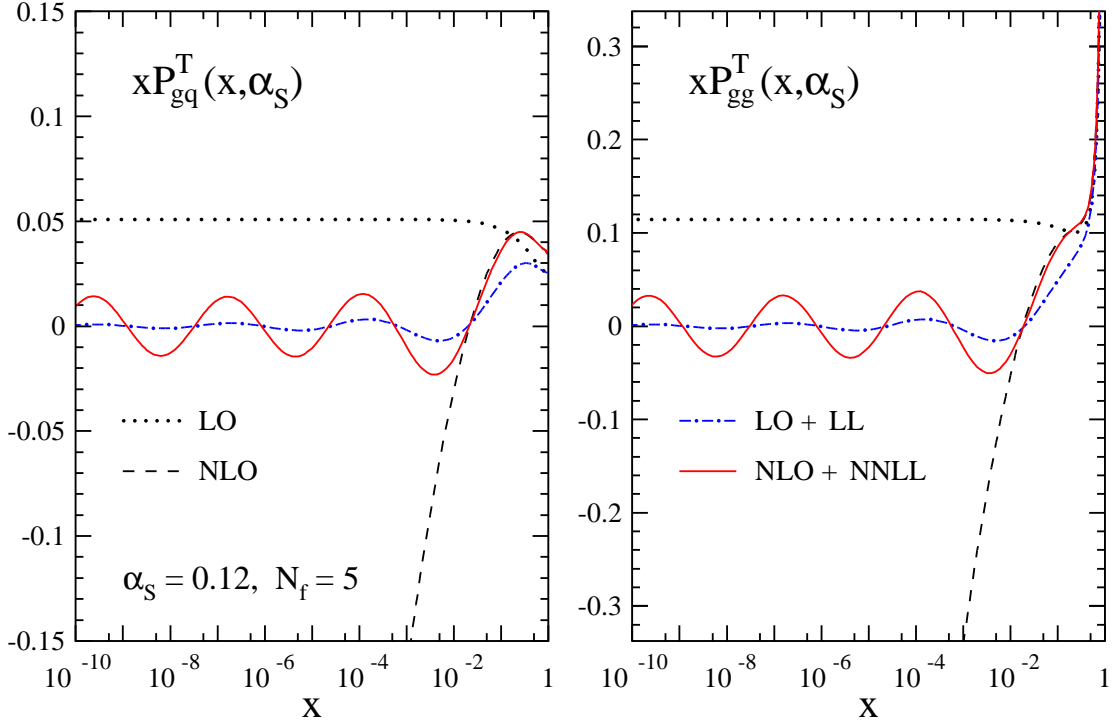,width=14.8cm,angle=0}\quad}
\vspace{-3mm}
\caption{ \label{fig:Pgitres}
The timelike gluon-quark and gluon-gluon splitting functions (multiplied by 
$x\,$), shown for a very wide range of the momentum fraction $x$ at a typical 
value of the strong coupling constant $\as$.
The all-$x$ (minimal $N\!=\!1$ finite) `LO$\,+\,$resummed' and 
`NLO$\,+\,$resummed' approximations are compared to the corresponding LO and 
NLO results valid only at large $x$, e.g., $x \protect\gsim 10^{\,-2}$ for 
NLO.}
\vspace{-2mm}
\end{figure}

\begin{figure}[p]
\vspace{-1mm}
\centerline{\epsfig{file=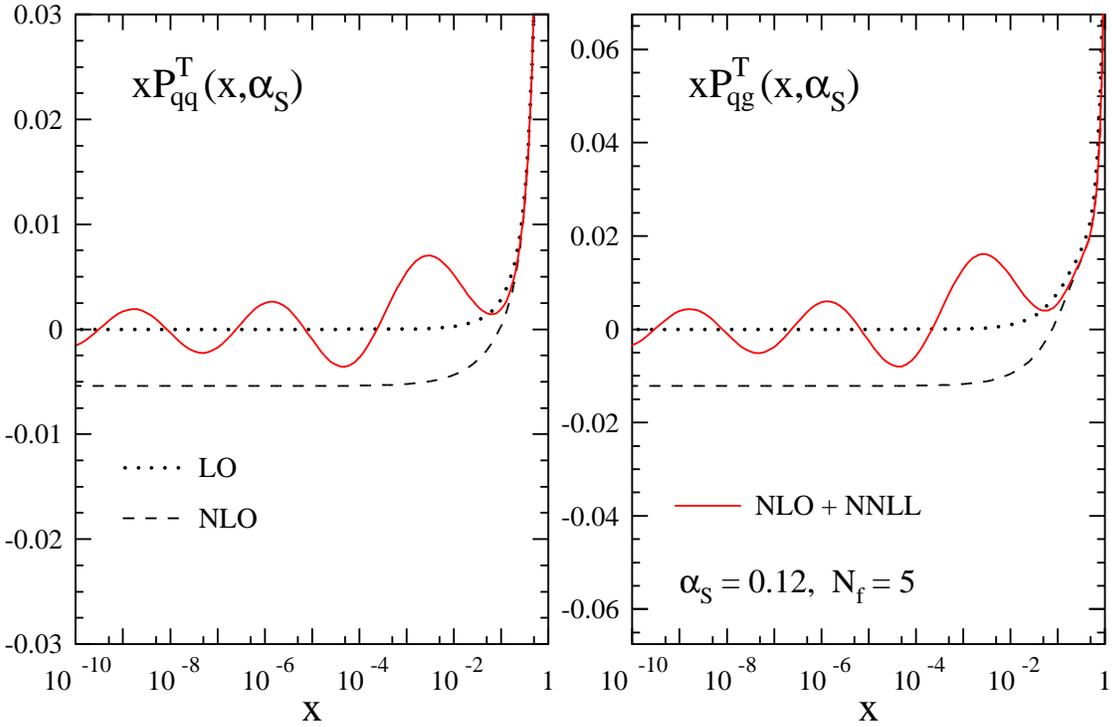,width=14.8cm,angle=0}\quad}
\vspace{-3mm}
\caption{ \label{fig:Pqitres}
 As Fig.~\ref{fig:Pgitres}, but for the timelike quark-quark and quark-gluon 
 splitting functions, for which the LO contributions does not include $1/x$ 
 terms and the resummation starts at NLL level.}
\vspace{-2mm}
\end{figure}

The LO$\:+\:$LL and NLO$\:+\:$NNLL approximations, which provide the
minimal $N\!=\!1$ finite resummations of the respective fixed-order results,
are illustrated in Fig.~\ref{fig:Pgitres} for the second column and in 
Fig.~\ref{fig:Pqitres} for the first column of the splitting-function matrix 
$P$ in Eq.~(\ref{CPmat}). 
Note that the scales of the right and left parts of both figures are related 
by a factor $\,\ca/\cf = 9/4$. The close similarity of the curves at small $x$
thus directly demonstrates the approximate `Casimir scaling' of 
$P_{\!\rm gi}^{\,T}$ and $P_{\!\rm qi}^{\,T}$.
The oscillation amplitudes of the former quantities are much larger than 
those of the latter. This behaviour is not due to the LL contributions to
$P_{\!\rm gi}^{\,T}$ which, as for other resummations, are numerically small.
There is no reason to assume that terms beyond the present NNLL
accuracy will be small. The Mellin inversion of Eqs.~(\ref{Pgg-n3}) and
(\ref{Pgg-n4}), however, appears to indicate that the amplitudes will
stabilize, for most of the wide $x$-range shown in the figures, after 
including the N$^3$LL and N$^4$LL contributions.

\begin{figure}[p]
\vspace*{-1mm}
\centerline{\epsfig{file=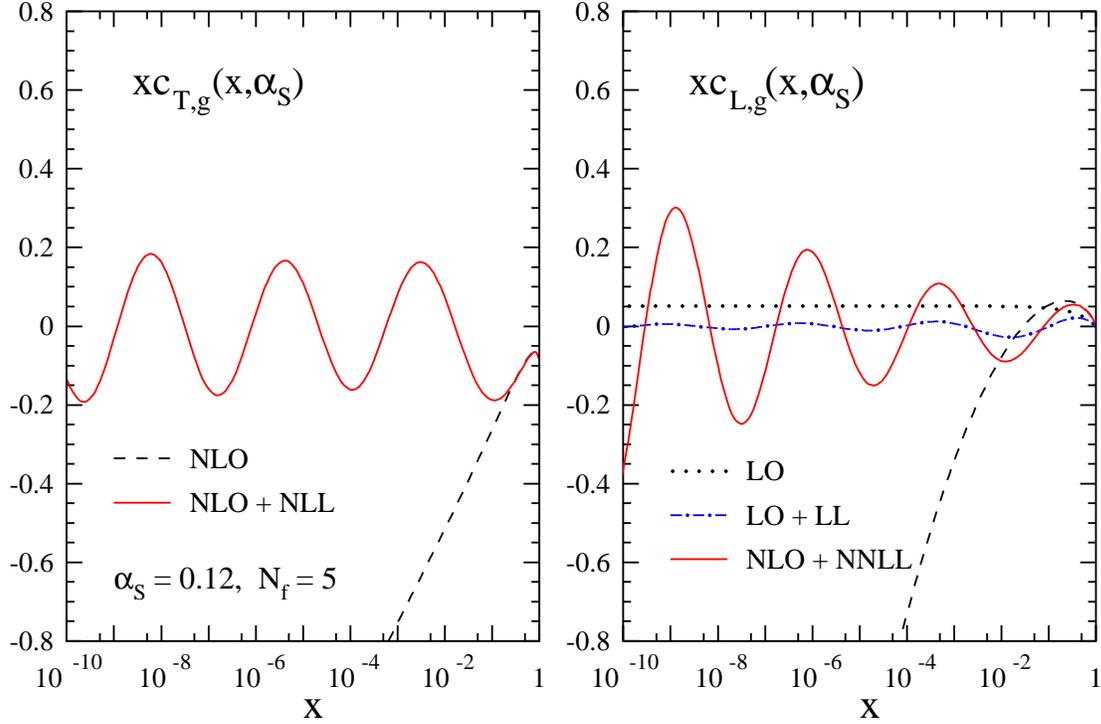,width=14.8cm,angle=0}\quad}
\vspace{-3mm}
\caption{ \label{fig:cagtres}
The all-$x$ gluon coefficient functions for the fragmentation functions $F_T$
and $F_L$ (multiplied by $x\,$), down to extremely small values of the
scaling variable $x$. As discussed above Eqs.~(\ref{cTg-cl}) and 
(\ref{AC-KTL}), the respective `(N)LO$\,+\,$resummed' approximations are 
defined by resumming as many logarithms as required to remove all $1/x$ terms 
due to the (N)LO contributions.}
\vspace{-2mm}
\end{figure}

\begin{figure}[p]
\vspace{-1mm}
\centerline{\epsfig{file=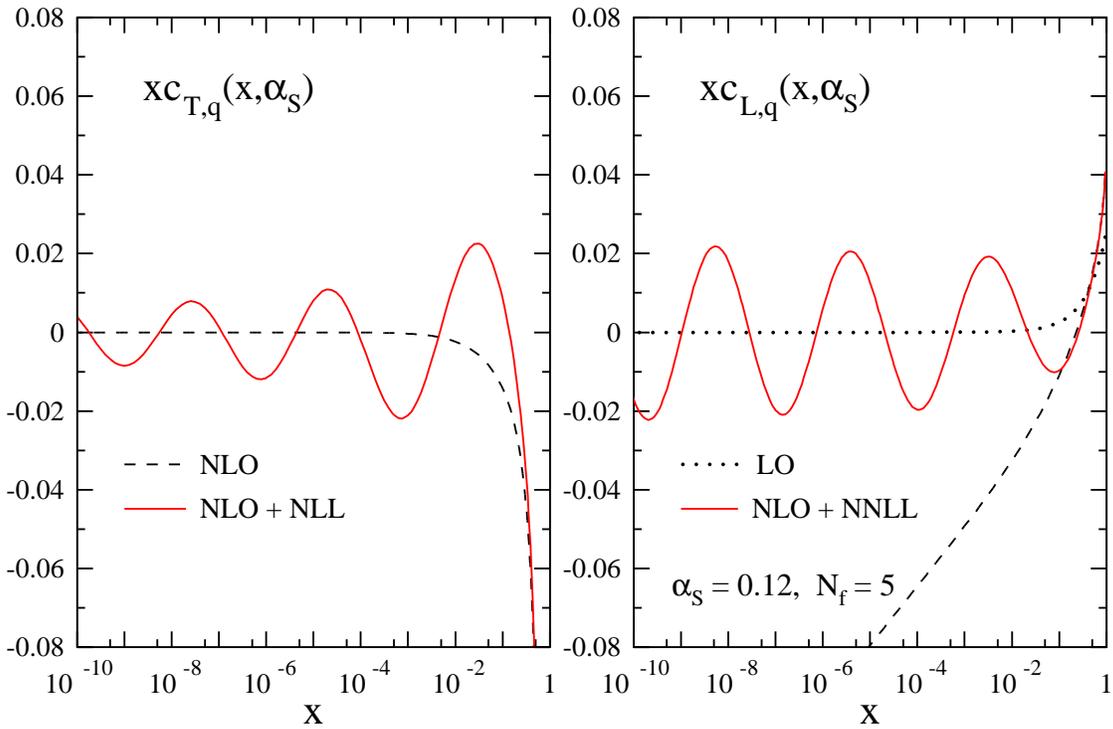,width=14.8cm,angle=0}\quad}
\vspace{-3mm}
\caption{ \label{fig:caqtres}
 As Fig.~\ref{fig:cagtres}, but for the quark coefficient functions which 
 are suppressed by one power of $\ln x\,$ w.r.t.\ the gluon quantities. 
 As the diagonal splitting functions, $C_{T,\rm q}$ includes a $\delta(1-x)$ 
 term.}
\vspace{-2mm}
\end{figure}

Corresponding minimal `LO$\:+\:$resummed' and `NLO$\:+\:$resummed' $x$-space
results are shown in Figs.~\ref{fig:cagtres} and \ref{fig:caqtres} for the
coefficient functions for $F_T$ (left) and $F_L$ (right). 
In this counting, the leading 
contribution $\cf/\ca\:c^{}_{\rm LL}$ to $C_{T,\rm g}$ in Eq.~(\ref{cTg-cl}) 
is not combined with the $\delta(1\!-\!x)$ LO term in Eq.~(\ref{Cexp}). 
This is consistent with the fact that $c^{}_{\rm LL}$ is a scheme-dependent 
quantity as discussed around Eq.~(\ref{ZsLL}).
Given the $\as$-expanded N$^3$LL result in Ref.~\cite{AV2011}, we expect that 
the coefficient function $C_{T,\rm q}$, for which Fig.~\ref{fig:caqtres} 
includes the Mellin inverse of Eq.~(\ref{cTq-cl}), will be the first quantity 
determined at an $N\!=\!1$ finite `NNLO$\:+\:$resummed' accuracy. 
The opposite is true for $C_{L,\rm g}$, the least stable quantity at the 
present accuracy as shown in the right part of Fig.~\ref{fig:cagtres}. Here,
as for $P_{\rm gi}^{\,T}$, the highest five logarithms need to be 
resummed for an all-$x$ combination with the NNLO~results.



\section{Summary}

\vspace*{-2mm}
We have presented the analytic all-order expressions in Mellin-$N$ space for 
the next-to-next-to-leading logarithmic (NNLL) small-$x$ resummation of the 
splitting functions $P_{\!ji}^{\,T}$ for parton fragmentation and the 
coefficient functions for the fragmentation functions $F_T$, $F_L$ and $F_\phi$
in gauge-boson and (heavy-top) Higgs-exchange semi-inclusive $e^+e^-$ 
annihilation.
The resummation replaces the double-logarithmic $x^{\,-1} \ln^{\,n\!} x$
small-$x$ enhancement of the fixed-order results for all these quantities by
an oscillatory behaviour which can be described in terms of Bessel functions 
for the splitting functions $P_{\!ji}^{\,T}(x,\as)$.
The present results are sufficient to construct a combined `NLO$\:+\:$resummed'
approximations which is applicable down to extremely small values of $x$ and 
determine the first three terms in the expansion of the first moments related 
to particle multiplicities in powers of~$\sqrt{\as}$.

In view of the dependence of the oscillation amplitudes on the logarithmic
order and the large size of the $N\!=\!1$ expansion parameter, an extension
of the present results to the fifth (N$^4L$) logarithms and a 
`NNLO$\:+\:$resummed' approximation is desirable. 
Completing these results is well beyond our present approach based on the
$D$-dimensional structure of the unfactorized fragmentation functions and the 
mass-factorization relations. However, we have taken a first step in this 
direction by confirming the all-order relation between initial-state and 
final-state splitting functions suggested in Refs.~\cite{DMS05,March06} at the
NNLL level and applying it to predict the `non-singlet' $\cf = 0$ part of 
$P_{\!gg}^{\,T}$ at N$^4$LL accuracy.

A {\sc Form} file of our NNLL results presented in the Sections 3, 4 and 5
above and Eqs.~(\ref{cTq-nn}) -- (\ref{cPhiq-nn}) below can be obtained by 
downloading the source of this article from the {\tt arXiv} servers. 

\subsection*{Acknowledgements}

\vspace*{-2mm}
A.V. and K.Y. are grateful to the organizers for the invitation to and 
hospitality at the workshop `Loops and Legs in Quantum Field Theory', 
Wernigerode (Germany), April 2012. Without this meeting the vital contribution 
of K.Y. to this article, the derivation of Eqs.~(\ref{AqSol1}) and 
(\ref{AqSol2}), would not have occurred.
We have profited from the {\sc Form} efficiency improvements provided by
Jos Vermaseren during the preparation of Ref.~\cite{AV2011}.
This research has been supported by the UK Science \& Technology Facilities
Council (STFC) under grant number ST/G00062X/1, the European-Union funded 
network {\it LHCPhenoNet} with contract number PITN-GA-2010-264564 and a
Discovery Grant of the Natural Sciences and Engineering Research Council
(NSERC) of~Canada.


\newpage
\renewcommand{\theequation}{A.\arabic{equation}}
\setcounter{equation}{0}
\section*{Appendix A}

\vspace*{-2mm}
The crucial step towards obtaining the analytic expressions presented above and 
in Appendix~B was the solution of the sequence $A_{qi}^{\,(n)}$ in Ref.\cite
{AV2011}. The key to understanding this sequence was the observation that
the denominators began as, and subsequently were divisors of, the sequence of 
least common multiples of triangular numbers \cite{A025555}. This suggested 
expanding each term as a sum of reciprocals of triangular numbers, which after
some `playing' with the initial terms led to  
\bea
\label{AqSol1}
  A_{qi}^{\,(2)} \!&\!=\!& \quad 1\;\; \;=\;\; \frct{1}{1} \; , 
\nn \\
  A_{qi}^{\,(3)} \!&\!=\!& \;\;\;\frct{11}{3}\; \;=\;\; 
   \frct{2}{3} \:+\: \frct{3}{1}  \; , 
\nn \\
  A_{qi}^{\,(4)} \!&\!=\!& \;\;\;\frct{73}{6}\; \;=\;\; 
   \frct{5}{6} \:+\: \frct{7}{3} \:+\: \frct{9}{1} \; , 
\nn \\  
  A_{qi}^{\,(5)} \!&\!=\!& \frct{1207}{30} \;=\; 
   \frct{14}{10} \:+\: \frct{19}{6} \:+\: \frct{23}{3} \:+\: \frct{28}{1} \; , 
\nn \\
  A_{qi}^{\,(6)} \!&\!=\!& \frct{2015}{15} \;=\;
   \frct{42}{15} \:+\: \frct{56}{10} \:+\: \frct{66}{6} \:+\: \frct{76}{3}
    \:+\: \frct{90}{1}  \; , \;\;\dots 
\eea
and the recognition that the successive numerators are the coefficients of the
series \cite{A028378} which are some particular sums of products of Catalan
numbers. Our knowledge of  $A_{qi}^{\,(n)}$ to $n = 17$ checks this 
identification to the 136$^{\,\rm th}$ entries in the sequences \cite{A025555} 
and \cite{A028378}, virtually excluding an agreement by coincidence. 
Using {\sc Maple}, this result was rewritten in the more accessible form
\bea
\label{AqSol2}
  A_{qi}^{\,(n)} \!&\!=\!&
   \frct{2 (2\,n-2)!}{(n-1)!(n+1)!}\, 
   \left( \frct{1}{n-1} + \frct{1}{n} + \frct{6}{n+1}  - 2 \right)
   \,+\,  \frct{2 (2\,n)!}{n!(n+1)!}\, \sum_{k\,=\,n}^{2\,n-3} \frct{1}{k} \; ,
\eea
which was then understood to represent the expansion coefficients [in terms
of $\xi$ in Eq.~(\ref{SLdefs})] of the first line in Eq.~(\ref{Pqq-cl}).
Once this function (the simplicity of which further supports the correctness
of the above deductions) was known, it was not too difficult to derive all
of Eqs.~(\ref{Pqq-cl}) -- (\ref{Pgq-cl}).

The corresponding $x$-space series can be readily obtained using 
Eqs.~(\ref{AqSol2}) and (\ref{Mlogs}). We have not been able to express the 
result as a known special function. It may be interesting for future studies, 
however, that the function $\cal L$ in Eq.~(\ref{SLdefs}) has a simple Mellin
(and Laplace) inverse, viz
\beq
\label{IMLog}
  \int_0^1\! dx \: x^{\,N-2}\, \: \frct{1}{\ln x}
  \left( J_0 (2\sqrt{a}\ln x) -1 \right) \:\;=\:\;
  \ln \Big( \frct{1}{2} + \frct{1}{2} \sqrt{1+ \frct{4\:\!a}{(N\!-\!1)^2} } \;
  \Big)
\eeq
with, in our case, $a = 8\,\ca\,\ar$. 
In view of Eq.~(\ref{dPgg-x}) and the logarithmic $x$-space arguments, the 
Mellin inverse of the first line of Eq.~(\ref{Pqq-cl}) and other products
of the form $(N-1)^{n} (S^{\,z}-1) {\cal L}$ with $z = 1,\,-1,\,-3,\:\dots\:$, 
can hence be expressed as Laplace convolutions of Bessel functions.


\renewcommand{\theequation}{B.\arabic{equation}}
\setcounter{equation}{0}
\section*{Appendix B}

\vspace*{-2mm}
Finally we present various, mostly lengthy results beyond the (at $N\neq 1$)
main `NLO $+$ resummed' approximation. 
We start with the N$^3$LL corrections to the splitting functions 
$P_{\,\rm qq}^{\,T}$ and $P_{\,\rm qg}^{\,T}$ given at NNLL accuracy in 
Eqs.~(\ref{Pqq-cl}) and (\ref{Pqg-cl}),

\pagebreak
\vspace*{-10mm}
\bea
\label{Pqq-n3}
  \lefteqn{P_{\,\rm qq\,,N^3LL}^{\,T}(N) \;=\;
    \frct{1}{2592}\*\frct{\cf\*\nf}{\cafo}\*\ar^2\*\Big\{
  - 3\*(121\*\cafo + 44\*\cath\*\nf - 88\*\cas\*\cf\*\nf + 4\*\cas\*\nfs 
  - 16\*\ca\*\cf\*\nfs 
  }\nn\\[-1mm]&&
  + 16\*\cfs\*\nfs)\*\frct{1}{\xi^2}\*
    ((S^{\,-2}-1-4\*\xi-16\*\xi^2)-(S^{\,-3}-1-6\*\xi)\*{\cal L})
  + 6\*(341\*\cafo + 40\*\cath\*\nf + 8\*\cas\*\cf\*\nf 
  \nn\\[-0.5mm]&&
  - 4\*\cas\*\nfs + 32\*\ca\*\cf\*\nfs - 48\*\cfs\*\nfs)\*
    \frct{1}{\xi^2}\*(S^{\,-1}-1-2\*\xi-6\*\xi^2)
  + 2\*([905 + 4608\*\,\z2]\*\cafo + 3456\*\cath\*\cf 
  \nn\\[-0.5mm]&&
  - 1576\*\cath\*\nf + 8680\*\cas\*\cf\*\nf + 556\*\cas\*\nfs 
    - 2336\*\ca\*\cf\*\nfs + 2064\*\cfs\*\nfs)\*
    \frct{1}{\xi^2}\*(S-1+2\*\xi+2\*\xi^2)
  \nn\\[-0.5mm]&&
  + 3\*([709 - 576\*\,\z2]\*\cafo - 300\*\cath\*\nf + 936\*\cas\*\cf\*\nf
    - 76\*\cas\*\nfs + 400\*\ca\*\cf\*\nfs - 496\*\cfs\*\nfs)\*
    \frct{1}{\xi^2}\*(S^{\,-1}
  \nn\\[-0.5mm]&&
  -1-2\*\xi)\*{\cal L}
  - 24\*(55\*\cafo - 12\*\cath\*\nf + 46\*\cas\*\cf\*\nf - 4\*\cas\*\nfs 
    + 20\*\ca\*\cf\*\nfs - 24\*\cfs\*\nfs)\*\frct{1}{\xi^2}\*(S^{\,-1}
  \nn\\[-0.5mm]&&
  -1-2\*\xi)\*{\cal L}^2
  - 6\*([5231 - 6048\*\,\z2]\*\cafo - [5184 - 6912\*\,\z2]\*\cath\*\cf 
    - 896\*\cath\*\nf - 2488\*\cas\*\cf\*\nf
  \nn\\[-0.5mm]&&
  - 292\*\cas\*\nfs + 2176\*\ca\*\cf\*\nfs - 3440\*\cfs\*\nfs)\*
     \frct{1}{\xi^2}\*(S-1+2\*\xi)\*{\cal L}
  - 24\*([961 - 864\*\,\z2]\*\cafo +  [1152\*\,\z2
  \nn\\[-0.5mm]&&
  - 612]\*\cath\*\cf - 336\*\cath\*\nf + 386\*\cas\*\cf\*\nf + 4\*\cas\*\nfs 
    + 92\*\ca\*\cf\*\nfs - 232\*\cfs\*\nfs)\*\frct{1}{\xi^2}\*
    (S-1+2\*\xi)\*{\cal L}^2
  \nn\\[-0.5mm]&&
  - 64\*([6 + 18\*\,\z2]\*\cafo + 27\*\cath\*\cf - 29\*\cath\*\nf 
  + 100\*\cas\*\cf\*\nf + 8\*\cas\*\nfs - 44\*\ca\*\cf\*\nfs + 60\*\cfs\*\nfs)\*
  \nn\\[-0.5mm]&&
  \,\cdot\, \frct{1}{\xi^2}\*(S-1)\*{\cal L}^3
  + 192\*([175 - 288\*\,\z2]\*\cafo - [261 - 288\*\,\z2]\*\cath\*\cf 
    + 41\*\cath\*\nf - 318\*\cas\*\cf\*\nf
  \nn\\[-0.5mm]&&
  - 28\*\cas\*\nfs + 176\*\ca\*\cf\*\nfs - 256\*\cfs\*\nfs)\*
    \frct{1}{\xi}\*{\cal L}^2
  + 8\*([3659 - 9216\*\,\z2]\*\cafo - [5832 - 8640\*\,\z2]\*\cath\*\cf
  \nn\\[-0.5mm]&&
  - 628\*\cath\*\nf - 3488\*\cas\*\cf\*\nf - 380\*\cas\*\nfs 
    + 2560\*\ca\*\cf\*\nfs - 3696\*\cfs\*\nfs)\*\frct{1}{\xi}\*(S-1+2\*\xi)
  \nn\\[-0.5mm]&&
  + 12\*([2017 + 576\*\,\z2]\*\cafo + 1728\*\cath\*\cf - 1772\*\cath\*\nf 
    + 5432\*\cas\*\cf\*\nf + 356\*\cas\*\nfs
  \nn\\[-0.5mm]&&
  - 1744\*\ca\*\cf\*\nfs + 2064\*\cfs\*\nfs)\*\frct{1}{\xi}\*(S-1)\*{\cal L}
  - 384\*([117 - 252\*\,\z2]\*\cafo - [207 - 288\*\,\z2]\*\cath\*\cf
  \nn\\[-1mm]&&
  + 6\*\cath\*\nf - 164\*\cas\*\cf\*\nf - 14\*\cas\*\nfs + 96\*\ca\*\cf\*\nfs 
    - 144\*\cfs\*\nfs)\*(1+\frct{1}{\xi}\*{\cal L})
  \Big\}
\eea

\vspace*{-4mm}
\nin
and

\vspace*{-6mm}
\bea
\label{Pqg-n3}
  \lefteqn{P_{\,\rm qg\,,N^3LL}^{\,T}(N) \,=\; 
  \frct{\ca}{\cf}P_{\,\rm qq\,,N^3LL}^{\,T}(N) + 
  \frct{1}{54}\*\frct{\nf}{\cath}\*\ar^2\*\bigg\{       
   2\*([81 - 144\*\,\z2]\*\cafo - 90\*\cath\*\cf + 144\*\,\z2\*\cath\*\cf 
   }\nn\\[-1mm]&&
  - 79\*\cath\*\nf + 106\*\cas\*\cf\*\nf + 6\*\cas\*\nfs - 24\*\ca\*\cf\*\nfs 
    + 24\*\cfs\*\nfs)\*\frct{1}{\xi^2}\*(S-1+2\*\xi+2\*\xi^2)
  \nn\\[-0.5mm]&&
  - 2\*([483 - 576\*\,\z2]\*\cafo 
  - [360 - 576\*\,\z2]\*\cath\*\cf - 139\*\cath\*\nf + 102\*\cas\*\cf\*\nf 
    + 56\*\ca\*\cf\*\nfs
  \nn\\&&
  - 6\*\cas\*\nfs - 88\*\cfs\*\nfs)\*\frct{1}{\xi}\*(S-1+2\*\xi)
  + ([429 - 576\*\,\z2]\*\cafo - [360 - 576\*\,\z2]\*\cath\*\cf 
  - 213\*\cath\*\nf 
  \nn\\[-1mm]&&
  + 250\*\cas\*\cf\*\nf - 2\*\cas\*\nfs + 40\*\ca\*\cf\*\nfs 
    - 72\*\cfs\*\nfs)\*\frct{1}{\xi^2}\*(S-1+2\*\xi)\*{\cal L}
  + 8\*([137 - 144\*\,\z2]\*\cafo 
  \nn\\[-1mm]&&
  - [90 - 144\*\,\z2]\*\cath\*\cf
  - 17\*\cath\*\nf - 6\*\cas\*\cf\*\nf - 6\*\cas\*\nfs + 36\*\ca\*\cf\*\nfs 
    - 48\*\cfs\*\nfs)\*(1+\frct{1}{\xi}\*{\cal L})
  \nn\\[-1mm]&&
  + (11\*\cafo + 13\*\cath\*\nf 
  - 26\*\cas\*\cf\*\nf + 2\*\cas\*\nfs - 8\*\ca\*\cf\*\nfs + 8\*\cfs\*\nfs)\*
    \frct{1}{\xi^2}\*(S^{\,-1}-1-2\*\xi)\*{\cal L}
  \nn\\[-1mm]&&
  + 4\*([59 - 72\*\,\z2]\*\cafo 
  - [45 - 72\*\,\z2]\*\cath\*\cf - 20\*\cath\*\nf + 20\*\cas\*\cf\*\nf 
    - 2\*\cas\*\nfs + 14\*\ca\*\cf\*\nfs 
  \nn\\[-1mm]&&
  - 20\*\cfs\*\nfs) \*\frct{1}{\xi^2}\*(S-1)\*{\cal L}^2 
  \bigg\}
\; .
\eea
Together with the first moments of the corresponding truncated second-order 
splitting functions 
\bea
\label{Pqq1N1}
  \overline{P}_{\,\rm qq}^{\,T(1)}(N\!=\!1) \!&\!=\!&
  - \left( 13 - 12\,\*\z2 + 8\,\*\z3 \right)\* \ca\*\cf
  + \left( 26 -24\,\*\z2 + 16\,\*\z3 \right)\* \cfs 
  + \frct{634}{27}\,\*\cf\*\nf
\; , \\[1mm]
\label{Pqg1N1}
  \overline{P}_{\,\rm qg}^{\,T(1)}(N\!=\!1) \!&\!=\!&
  \left( \frct{796}{27} - \frct{16}{3}\,\*\z2 \right) \*\ca\*\nf 
  + 4\,\*\cf\*\nf \; ,
\eea
which can be derived, e.g., using {\sc Form} packages \cite{FORM} based on
Refs.~\cite{Hsums,HPLs} as discussed in Ref.~\cite{AV2011}, 
Eqs.~(\ref{Pqq-n3}) and (\ref{Pqg-n3}) lead to the combined results
\bea
\label{Pqq-n3-N1}
 \left. P_{\,\rm qq}^{\,T}(N\!=\!1)\right|_{\ars} \!&\!=\!& 
  \frct{1}{162}\,\*\frct{\cf}{\cafo}\,\*
  \Big\{ -\,[2106 - 1944\,\*\z2 + 1296\,\*\z3]\,\*\cafi + [4477 - 2016\,\*\z2]
  \,\*\cafo\*\nf 
\nn\\&& \mbox{}
  + [4212 - 3888\,\*\z2 + 2592\,\*\z3]\,\*\cafo\*\cf 
  - 1124\,\*\cath\*\nfs + 1728\,\*\z2\,\*\cath\*\cf\*\nf 
\nn\\[-0.5mm]&& \mbox{}
  + 116\,\*\cas\*\nft + 2336\,\*\cas\*\cf\*\nfs 
  - 448\,\*\ca\*\cf\*\nft + 432\,\*\cfs\*\nft \Big\}
\; ,
\\[1mm]
\label{Pqg-n3-N1}
 \left. P_{\,\rm qg}^{\,T}(N\!=\!1)\right|_{\ars} \!&\!=\!&
  \frct{1}{162}\,\*\frct{\nf}{\cath}\,\*
  \Big\{ [3913 - 1152\,\*\z2]\,\*\cafo + 1728\,\*\cath\*\cf 
  - 812\,\*\cath\*\nf + 116\,\*\cas\*\nfs 
\nn\\[-1mm]&&
  + 2240\,\*\cas\*\cf\*\nf - 544\,\*\ca\*\cf\*\nfs + 624\,\*\cfs\*\nfs \Big\}
\; .
\eea
The numerical values for $\nf=5$ of these coefficients have been included in 
Eq.~(\ref{Pres-n1}) above.

The NNLL corrections to the coefficient functions (\ref{cTg-cl}) -- 
(\ref{cPq-cl}) for $F_T$ and $F_\phi$ can be written as
\bea
\label{cTq-nn}
  \lefteqn{C_{T,\rm q,\,NNL}^{}(N) \;=\; 
 \frct{1}{432}\*\frct{\cf\*\nf}{\cath}\*\ar\*\Big\{ - 192\*(5\,\*\cas 
  - 4\,\*\ca\*\nf + 12\,\*\cf\*\nf) \*(\frct{1}{\xi}\*{\cal L}+1)
  + 8\*(35\,\*\cas - 14\,\*\ca\*\nf 
}\nn\\[-0.5mm] \mbox{}
 & &
  + 24\,\*\cf\*\nf)\*\frct{1}{\xi}\*(F^{-3}-1+3\*\xi)
  - 8\,\*(35\,\*\cas + 34\,\*\ca\*\nf - 120\,\*\cf\*\nf)\*
    \frct{1}{\xi}\*(F^{-1}-1+\xi)
  - 3\,\*(505\,\*\cas 
\nn\\[-0.5mm] && \mbox{}
  - 150\,\*\ca\*\nf + 416\,\*\cf\*\nf)\*\frct{1}{\xi}\*(F^{-1}-1)\*{\cal L}
  - 4\,\*(887\,\*\cas - 122\,\*\ca\*\nf + 336\,\*\cf\*\nf)\*(F-1)
\nn\\[-0.5mm] && \mbox{}
  + 4\,\*(169\,\*\cas + 86\,\*\ca\*\nf - 276\,\*\cf\*\nf)\*
    \frct{1}{\xi}\*(F-1)\*{\cal L}
  + 12\,\*(11\,\*\cas + 2\,\*\ca\*\nf - 4\,\*\cf\*\nf)\*(4\,\*(F^{3}-1) 
\nn\\[-0.5mm] && \mbox{}
  - \frct{1}{\xi}\*(F^{5}-1)\*{\cal L})
  + (11\,\*\cas - 2\,\*\ca\*\nf)\*(20\,\*(F^{5}-1) 
  + 6\,\*\frct{1}{\xi}\*(F^{3}-1)\*{\cal L}-5\,\*
   \frct{1}{\xi}\*(F^{7}-1)\*{\cal L})
\nn\\[-0.5mm] && \mbox{}
  + 96\,\*(5\,\*\cas - 2\,\*\ca\*\nf + 6\,\*\cf\*\nf)\*
    \frct{1}{\xi}\*F\*{\cal L}^2
  \Big\}
%
\\[1mm]
\label{cTg-nn}
  \lefteqn{C_{T,\rm g,\,NNL}^{}(N) = -\frct{\cf}{\ca}C_{T\,\rm q,\,NNL}^{}(N) 
    + \frct{1}{82944}\*\frct{\cf}{\cafo}\*\ar\*\Big\{ 
   9216\,\*( [ 54 - 72\,\*\z2 ]\* \cafo - [ 45 - 72\*\,\z2]\*\cath\*\cf
}\nn\\[-0.5mm] \mbox{}
 & &
  - 23\,\*\cath\*\nf + 20\,\*\cas\*\cf\*\nf)\*(1-\frct{1}{\xi}\*
    (S-1+2\*\xi-{\cal L}))
  - 36864\,\*(\cas\*\cf\*\nf + \ca\*\cf\*\nfs 
\nn\\[-0.5mm] && \mbox{}
  - 2\,\*\cfs\*\nfs)
  \*\frct{1}{\xi}\*((S-1+2\*\xi)+(F-1)\*{\cal L})
  + 256\,\*( [ 637 - 1620\,\*\z2 ] \*\cafo - [ 810 - 1296\*\,\z2 ] \*\cath\*\cf
\nn\\[-0.5mm] && \mbox{}
  - 206\,\*\cath\*\nf + 129\,\*\cas\*\cf\*\nf - 44\,\*\cas\*\nfs 
  + 258\,\*\ca\*\cf\*\nfs - 342\,\*\cfs\*\nfs)\*\frct{1}{\xi}\*(F^{-3}-1+3\*\xi)
\nn\\[-0.5mm] && \mbox{}
  + 256\,\*( [ 1523 - 972\,\*\z2 ] \*\cafo - [ 810 - 1296\*\,\z2 ] \*\cath\*\cf   - 262\,\*\cath\*\nf + 159\,\*\cas\*\cf\*\nf - 28\,\*\cas\*\nfs 
\nn\\ && \mbox{}
  + 318\,\*\ca\*\cf\*\nfs - 522\,\*\cfs\*\nfs)\*\frct{1}{\xi}\*(F^{-1}-1+\xi)
  + ( [ 1285825 - 1327104\,\*\z2 ] \*\cafo - 331776\,\* 
\nn\\ && \,\cdot\,
  [1 - 2\*\,\z2 ] \*\cath\*\cf - 24716\,\*\cath\*\nf - 19392\,\*\cas\*\cf\*\nf 
   - 45020\,\*\cas\*\nfs + 233088\,\*\ca\*\cf\*\nfs 
\nn\\[0.5mm] && \mbox{}
  - 290304\,\*\cfs\*\nfs)\*(F-1)
  + 8\,\*(14839\,\*\cafo + 1168\,\*\cath\*\nf - 2172\,\*\cas\*\cf\*\nf 
  - 284\,\*\cas\*\nfs 
\nn\\[0.5mm] && \mbox{}
  + 1128\,\*\ca\*\cf\*\nfs - 1152\,\*\cfs\*\nfs)\*(F^{3}-1)
  + ( [ 75461 + 165888\,\*\z2 ] \*\cafo + 41156\,\*\cath\*\nf 
\nn\\[0.5mm] && \mbox{}
  - 110016\,\*\cas\*\cf\*\nf 
  + 11444\,\*\cas\*\nfs - 54144\,\*\ca\*\cf\*\nfs 
  + 62208\,\*\cfs\*\nfs)\*(F^{5}-1)
  + 16\,\*(11957\,\*\cafo 
\nn\\[0.5mm] && \mbox{}
  - 40\,\*\cath\*\nf - 4488\,\*\cas\*\cf\*\nf 
  - 388\,\*\cas\*\nfs + 816\,\*\ca\*\cf\*\nfs)\*(F^{7}-1)
  - (21901\,\*\cafo + 55396\,\*\cath\*\nf 
\nn\\[0.5mm] && \mbox{}
  - 63360\,\*\cas\*\cf\*\nf + 724\,\*\cas\*\nfs 
  - 11520\,\*\ca\*\cf\*\nfs + 11520\,\*\cfs\*\nfs)\*(F^{9}-1)
  - 840\,\*(121\,\*\cafo 
\nn\\ && \mbox{}
  - 44\,\*\cas\*\cf\*\nf - 4\,\*\cas\*\nfs
  + 8\,\*\ca\*\cf\*\nfs)\*(F^{11}-1)
  - 385\,\*\cas\*(11\,\*\ca - 2\,\*\nf)^2\*(F^{13}-1)
  \Big\}
\eea
and
\bea
\label{cPhig-nn}
  \lefteqn{C_{\phi,\rm g,\,NNL}^{}(N) \;=\; -C_{T\,\rm q,\,NNL}^{}(N) \:+\:
  \frct{1}{82944}\,\*\frct{1}{\cath}\,\*\ar\,\*\Big\{
  256\,\*( [ 601 - 324\*\,\z2 ]\*\cafo + 64\,\*\cath\*\nf 
  + 93\,\*\cas\*\cf\*\nf 
}\nn\\[-1mm] \mbox{}
 & &
  - 8\,\*\cas\*\nfs - 30\,\*\ca\*\cf\*\nfs + 90\,\*\cfs\*\nfs)\*
    \frct{1}{\xi}\*((S-1)\*F^{-1}+2\*\xi)
  - 18432\,\*(2\,\*\cas\*\cf\*\nf - \ca\*\cf\*\nfs 
\nn\\[-1mm]&& \mbox{}
  + 2\,\*\cfs\*\nfs)\*\frct{1}{\xi}\*(S-1)\*F\*{\cal L}
  + ( [ 3335233 - 663552\*\,\z2 ] \*\cafo - 162380\,\*\cath\*\nf 
    - 356928\,\*\cas\*\cf\*\nf 
\nn\\[-0.5mm]&& \mbox{}
  - 7004\,\*\cas\*\nfs + 46464\,\*\ca\*\cf\*\nfs - 69120\,\*\cfs\*\nfs)\*(F-1)
  + (9559\,\*\cafo + 3256\,\*\cath\*\nf + 244\,\*\cas\*\nfs
\nn\\[0.5mm]&& \mbox{}
  - 6732\,\*\cas\*\cf\*\nf - 1080\,\*\ca\*\cf\*\nfs 
  + 1152\,\*\cfs\*\nfs)\,\*8\,\*(F^{3}-1)
  - ( [ 55483 - 165888\*\,\z2 ] \*\cafo 
\nn\\[0.5mm]&& \mbox{}
  - 17540\,\*\cath\*\nf + 52032\,\*\cas\*\cf\*\nf - 4148\,\*\cas\*\nfs 
  + 30336\,\*\ca\*\cf\*\nfs - 43776\,\*\cfs\*\nfs)\*(F^{5}-1)
\nn\\[0.5mm]&& \mbox{}
  + 16\,\*(9317\,\*\cafo - (220\,\*\cath + 3168\,\*\cas\*\cf) \*\nf 
  - (268\,\*\cas - 576\,\*\ca\*\cf) \*\nfs)\*(F^{7}-1)
  - (21901\,\*\cafo 
\nn\\[0.5mm]&& \mbox{}
  + 55396\,\*\cath\*\nf - 63360\,\*\cas\*\cf\*\nf + 724\,\*\cas\*\nfs 
  - 11520\,\*\cf\*\nfs\*(\ca - \cf))\*(F^{9}-1) 
  - (121\,\*\cafo 
\nn\\[-0.5mm]&& \mbox{}
  - 44\,\*\cas\*\cf\*\nf - 4\,\*\cas\*\nfs + 8\,\*\ca\*\cf\*\nfs)
    \,\*840\,\*(F^{11}-1)
  - 385\,\*(11\,\*\ca - 2\,\*\nf)^2\*\cas\*(F^{13}-1)\Big\}
%
\\[2mm]
\label{cPhiq-nn}
  \lefteqn{C_{\phi,\rm q,\,NNL}^{}(N) \;=\; 
  \frct{\ca}{\cf}\:\*C_{T\,\rm q,\,NNL}^{}(N) \:+\:
    \frct{2}{9}\,\*\frct{\nf}{\cas}\,\*\ar\,\*\Big\{ 
  - 2\,\*(\cas + \ca\*\nf - 2\,\*\cf\*\nf)\*(1+\frct{1}{\xi}\*{\cal L})
}\nn\\[-0.5mm] & & \mbox{}
  - (2\,\*\cas - \ca\*\nf + 2\,\*\cf\*\nf)\*     
     \frct{1}{\xi}\*\left((F^{-3}-1+3\,\*\xi)-(F^{-1}-1)\*{\cal L}\right)
\nn\\[-0.5mm]&& \mbox{}
  + (4\,\*\cas + \ca\*\nf - 2\,\*\cf\*\nf)\*
     \frct{1}{\xi}\*\left((F^{-1}-1+\xi)-(F-1)\*{\cal L}\right)\Big\}
\eea

Finally we write down the N$^4$LL contribution to the splitting function 
$P_{\rm gg}^{\,T}$ in the limit $\cf = 0$, which still contains one unknown 
coefficient as discussed above,
\bea
\label{Pgg-n4}
  \lefteqn{  \left. P_{\rm gg}^{\,T}(N) \right|_{\rm N^4LL}^{C_F=0} \;=\; 
  \frct{1}{13271040}\,\*\frct{1}{\cas}\,\*\ars\,\*\bar{N}\,\*
  \Big\{16\*([15688235 - 19918080\*\,\z2 + 7983360\*\,\z3 
  }\nn\\&& \mbox{}
  + 5059584\*\,\zs2]\*\cafo + [914360 + 875520\*\,\z2 
  - 2142720\*\,\z3]\*\cath\*\nf - [134200 + 46080\*\,\z2]\,\*\cas\*\nfs
\nn\\[0.5mm]&& \mbox{}
  + 5600\*\nft\*\ca - 80\*\nffo)\*(S-1)
  - 32\*([7822505 - 2826000\*\,\z2 + 1330560\*\,\z3 - 134784\*\,\zs2]\,\*\cafo
\nn\\[0.5mm]&& \mbox{}
  + [514490 + 83520\*\,\z2 - 276480\*\,\z3]\,\*\cath\*\nf 
  + [16880 + 20160\*\,\z2]\,\*\cas\*\nfs - 2840\,\*\nft\*\ca + 80\,\*\nffo)
\nn\\&& \,\cdot\,
  \*\frct{1}{\xi}\*(S-1+2\*\xi)
  + 2\*([12686895 + 12997440\*\,\z2 - 2471040\*\,\z3 
  - 10907136\*\,\zs2]\,\*\cafo
\nn\\&& \mbox{}
  - [3309880 + 564480\*\,\z2 - 1624320\*\,\z3]\,\*\cath\*\nf 
  + [37960 - 172800\*\,\z2]\*\cas\*\nfs + 21280\,\*\nft\*\ca
\nn\\&& \mbox{}
  - 1040\*\nffo)\*\frct{1}{\xi^2}\*(S-1+2\*\xi+2\*\xi^2)
  - 4\*([3135445 + 6822720\*\,\z2 + 190080\*\,\z3 - 5868288\*\,\zs2]\,\*\cafo
\nn\\&& \mbox{}
  - [1973120 + 587520\*\,\z2 - 1071360\*\,\z3]\,\*\cath\*\nf 
  + [106520 - 149760\*\,\z2]\,\*\cas\*\nfs + 14080\,\*\nft\*\ca
\nn\\&& \mbox{}
  - 1200\,\*\nffo)\*\frct{1}{\xi^2}\*(S^{\,-1}-1-2\*\xi-6\*\xi^2)
  - ([2095591 + 158976\*\,\z2 - 1140480\*\,\z3 + 331776\*\,\zs2]\,\*\cafo
\nn\\&& \mbox{}
  + [61560 + 396288\*\,\z2 - 207360\*\,\z3]\,\*\cath\*\nf 
  - [83352 - 64512\*\,\z2]\,\*\cas\*\nfs + 224\,\*\nft\*\ca + 880\,\*\nffo)
\nn\\&& \,\cdot\,
  \*5\,\*\*\frct{1}{\xi^2}\*(S^{\,-3}-1-6\*\xi-30\,\*\xi^2)
  - 10\*([198803 - 209088\*\,\z2]\,\*\cafo 
  + [69872 - 76032\*\,\z2]\,\*\cath\*\nf
\nn\\[-0.5mm]&& \mbox{}
  - [600 + 6912\*\,\z2]\,\*\cas\*\nfs - 2368\,\*\nft\*\ca 
  - 208\,\*\nffo)\*\frct{1}{\xi^2}\*(S^{\,-5}-1-10\,\*\xi-70\,\*\xi^2)
\nn\\[-0.5mm]&& \mbox{}
  - 25\*(11\,\*\ca+2\,\*\nf)^4\*\frct{1}{\xi^2}\* 
    (S^{\,-7}-1-14\,\*\xi-126\,\*\xi^2) 
  \Big\} \,+\, 
  \,\cas\*\,\ars\,\*\bar{N}\,\* (S+S^{\,-1}-2)\,\* 
  B_{\rm gg}^{\,S(3)} \; ,
\eea
where the normalization of the last coefficient has been given below 
Eq.~(\ref{Pggres-n1}). This result leads to the first moment
\bea
\label{Pgg-n4-N1}
 \left. P_{\rm gg}^{\,T}(N\!=\!1)\right|_{\,\ar^{5/2}}^{\,\cf=0} \!&\!=\!&
  (2\,\*\ca)^{1/2}\,\* \frct{1}{41472}\,\*\frct{1}{\cas}\,\*
  \Big\{ 
  \left[ 182872 + 175104\,\*\z2 - 428544\,\*\z3 \right]\* \cath\*\nf  
\nn\\&& \mbox{\hspn}
    +\, \left[ 3137647 - 3983616\,\*\z2 + 1596672\,\*\z3
    + 5059584/5\,\,\*\zs2 \right]\* \cafo
\\&& \mbox{\hspn}
    -\, \left[ 26840 + 9216\,\*\z2 \right]\,\*\cas\*\nfs 
    + 1120\,\*\ca\,\*\nft - 16\,\*\nffo
  \,+\, 165888\,\*\cafo\,\*  
        B_{\rm gg}^{\,S(3)}
  \Big\} \quad \nn
\eea
which has been included in an approximate numerical form for $\nf=5$
Eq.~(\ref{Pggres-n1}).



\end{document}